\documentstyle[12pt,epsfig]{article}
\newcommand{\beq}{\begin{equation}}
\newcommand{\eeq}{\end{equation}}
\newcommand{\bea}{\begin{eqnarray}}
\newcommand{\eea}{\end{eqnarray}}

\newcommand{\gsim}{\lower.7ex\hbox{$
\;\stackrel{\textstyle>}{\sim}\;$}}
\newcommand{\lsim}{\lower.7ex\hbox{$
\;\stackrel{\textstyle<}{\sim}\;$}}

\newcommand{\eod}{\end{document}}

\def\cp{{\bf CP}}
\def\cpt{{\bf CPT}}

\begin{document}
\thispagestyle{empty}
\vspace*{-22mm}

\vspace*{1.3mm}

\begin{center}
{\Large {\bf On a CP Anisotropy Measurement in the Dalitz Plot
}}


\vspace*{19mm}

{\bf I. Bediaga$^a$, I.I.~Bigi$^b$, A. Gomes$^a$, G. Guerrer$^a$, J. Miranda$^a$ 
and A.C. dos Reis$^a$}\\
\vspace{7mm}
$^a$ {\sl  Centro 
Brasileiro de Pesquisas F\'\i sicas, Rua Xavier Sigaud 150, 22290-180, Rio de Janeiro, RJ, Brazil}\\
$^b$ {\sl Department of Physics, University of Notre Dame du Lac}
\vspace*{-.8mm}\\
{\sl Notre Dame, IN 46556, USA}\\
{\sl email addresses: bediaga@cbpf.br, ibigi@nd.edu, guerrer@cbpf.br,
jussara@cbpf.br, alvaro@cbpf.br, alberto@cbpf.br} \\

\vspace*{10mm}

{\bf Abstract}\vspace*{-1.5mm}\\
\end{center}
We describe a novel use of the Dalitz plot to probe \cp~symmetry in three-body modes of 
$B$ and $D$ mesons. It is based on an observable inspired by astronomers' practice, namely the 
significance in the difference between corresponding Dalitz plot bins. It provides a 
model independent mapping of local \cp~asymmetries.  We illustrate the 
method for probing \cp~symmetry in the two complementary cases of 
$B$ and $D$ decays: in the former sizable or even large effects can be expected, yet have to be differentiated 
against leading Standard Model contributions, while in the latter one cannot count on sizable effects, 
yet has to deal with much less Standard Model background.

\vspace{3mm}

\hrule

\tableofcontents
\vspace{5mm}

\hrule\vspace{5mm}

\section{Prologue}

While the announcement of the 2008 Nobel Prize in Physics has made official the status of 
KM theory as the main source of \cp~violation as observed in $K$ and $B$ decays, it does not close 
the chapter on it for three main reasons \cite{CPBOOK}: 
\begin{itemize}
\item 
We know baryogenesis in our Universe requires New Physics with \cp~violation. 
\item 
A host of largely theoretical arguments suggests -- persuasively in our view -- that 
New Physics exists around the TeV scale with rich dynamical structures. In general those can provide 
several novel sources of \cp~violation. In that context one uses the high sensitivity of \cp~studies 
as a tool to search for New Physics and hopefully infer some of its salient features. One should keep in mind that 
observable \cp~asymmetries can be {\em linear} in a New Physics amplitude 
with the Standard Model (SM) providing the other one; therefore one  achieves sensitivity to small contributions. 
\item 
The LHCb experiment \cite{jinst} is poised to acquire large sets of high quality data on the decays of $B$ and $D$ mesons. 
\end{itemize}
The stage for \cp~studies has recently become wider with the observation of $D^0 - \bar D^0$ oscillations 
\cite{BABAROSC,BELLEOSC}. Fortunately we can expect to continue our quest for \cp~violation with the continuing work 
of the Belle Collaboration, the hoped for realization of a Super-B Factory \cite{VALENCIA} and in particular with the 
beginning of the LHCb experiment. 

We will focus on three-body final states in the decays of $B$ and $D$ mesons and a novel strategy 
to probe \cp~symmetry in their Dalitz plots.  Those two classes of transitions offer complementary challenges both 
on the experimental and theoretical side. 
\begin{itemize}
\item 
KM dynamics is expected to generate large \cp~asymmetries -- 
to the tune of, say, 10 - 20 \% -- in modes like $B \to \pi \pi\pi$, $K\pi\pi$, $Kp \bar p$. Since those command 
very small branching ratios, we still need very large samples of $B$ mesons, as will be produced by the 
LHCb experiment. While we can be confident that such effects will be observed  
(supporting the experimentalists' enthusiasm), the real challenge at that time will be how to interpret a  signal: 
does it reveal the presence of New Physics or is it consistent with being 
from KM dynamics alone? This will represent a  non-trivial conundrum, since all the available evidence points to 
New Physics mustering no more than {\em non-}leading contributions in $B$ decays; this is often referred to as 
the "Flavour Problem of New Physics". We will have a realistic chance to answer this question only if we have accurate 
as well as comprehensive data. 

\item 
On the other hand in most {\em charm} decays experimental bounds tell us 
we can hope for at best moderate size asymmetries even in the presence of New Physics. 
The redeeming feature is that the SM can generate \cp~violation at most at the 
$10^{-3}$ level in the `best' cases and even significantly less in others. However we view the latter as good news: 
for almost any asymmetry observed in the near future would represent strong evidence for the intervention of New Physics. 
Using the language of  ~`signal-to-noise'  ~familiar from experimental studies: while we expect significantly smaller 
asymmetries in $D$ than in $B$ decays, the 
{\em theoretical}  ~`noise' or `background' -- i.e. contributions from the SM -- is even more reduced in the former than 
the latter. Thus we conjecture 
\beq 
\left. \frac{\rm \cp~asymmetry}{\rm SM\; contribution}\right|_{{\bf D}\; {\rm decays}} > 
\left. \frac{\rm \cp~asymmetry}{\rm SM\; contribution}\right|_{{\bf B}\; {\rm decays}}
\eeq
The charm branching ratios are typically  sizable and we have already acquired a great deal of experience in describing 
them and their Dalitz plots. The central challenge then is to control systematics to a degree that allows probing 
asymmetries down to $10^{-2}$ or even better. 

\end{itemize}

So far no \cp~asymmetry has been {\em established} on the five sigma significance level. However we expect that to change soon and actually predict Dalitz studies to become one of the central tools for \cp~probes. 
\begin{itemize}
\item 
{\em Accurate} measurements of \cp~asymmetries will be a necessary (though probably not 
sufficient) condition for deciding whether they reveal the intervention of New Physics or not. In the case 
of charm decays the anticipated small size of effects constitutes the main challenge; in $B$ decays, on the other 
hand, we have to learn how to subtract the presumably leading SM contribution.  Dalitz plot descriptions with their many correlations yield 
{\em over}constraints, providing reliable  
validation tools.  Tracking the time evolution in 
$D^0$, $B_d$ and $B_s$ transitions characteristic of oscillations will further illuminate the underlying dynamics. 
\item 
Establishing the intervention of New Physics in \cp~studies will of course represent a seminal achievement, yet we want 
to do even better. For our goal has to be to infer salient features of that anticipated New Physics. Some of those 
can be read off the flavour structure of the final states. Yet to infer the Lorentz structure of the underlying operator 
we have to go beyond final states consisting of two pseudoscalar or 
one pseudoscalar and one vector meson; for their amplitudes are described by a single number. However, once we have 
three pseudoscalar mesons or a baryon-antibaryon pair plus a meson, the 
kinematics is no longer trivial, and final state distributions can tell us whether spin-zero or spin-one couplings 
are involved in the transition operator. This important feature will be illustrated below. 

\end{itemize}
Fortunately pioneering work has been done by Belle and BaBar: Based on a Dalitz plot 
study they have extended their probe of \cp~invariance to quasi-two-body channels involving 
resonances and indeed found intriguing {\em evidence} for a direct \cp~asymmetry in the  
mode $B^{\pm} \to K^{\pm}\rho^0$ \cite{BaBark2pi,bellek2pi,BaBark2pi2nd}, which -- if established -- would be a first. 

Relying on mass projections is only one way to use the dynamical information contained in a 
Dalitz plot, and it cannot be expected to harness the full potential of Dalitz studies. It is the 
{\em interference} between two 
{\em neighbouring resonances} that presumably provides the most sensitive \cp~probe. For  
a direct \cp~asymmetry to surface one needs the interplay of a weak and a strong phase with the former in contrast to the latter changing signs under a \cp~transformation. For the latter one usually takes the 
strong phase shifts,  which cannot be calculated from first principles. Yet when one deals with a finite-width 
resonance, its Breit-Wigner parameters can provide the required strong phase, which varies 
with the mass bin in a characteristic and largely predictable way. This feature can provide further 
validation for the experimental findings. 
Alas -- a full-fledged Dalitz plot description requires huge statistics and considerable theoretical `overhead' in 
selecting the transitions deemed relevant and parametrizing their amplitudes. It has to be the ultimate goal to 
develop such a complete description with as much accuracy as possible, yet that will be a long term task, and 
it is not clear what irreducible model dependence will remain. A full Dalitz plot  
description would help in the extraction of the CKM phases \cite{Helen,BGJ}. There are, however, some  theoretical 
issues concerning the large phase space of the B decay into three 
light mesons \cite{USP} that need to be understood.

To avoid such model dependence one can divide the Dalitz plot into bins, and then  directly compare the \cp~conjugate 
Dalitz plot regions in a bin-by-bin basis.
Yet results based on studying the ratio between the difference over the sum of the populations 
are quite vulnerable to fake effects from statistical fluctuations. Therefore we suggest a refinement of such a 
direct comparison, namely to study the {\em significance} of the difference. This proposal has been inspired by 
what has become standard routine in astronomy when analyzing light sources in the 
sky \cite{astro}. Its main values lies in three aspects:
\begin{itemize}
\item
As illustrated later it provides a model-independent and robust method to determine \cp~asymmetries already with limited 
statistics and identify the regions of a Dalitz plot, where they occur. 
\item 
This is particularly important when dealing with small or even tiny effects as expected in charm decays. 
\item 
Its findings provide powerful constraints on any full Dalitz plot model to emerge.  
\end{itemize} 
In talking about `limited' statistics we do {\em not} mean {\em small} statistics -- a situation addressed in 
\cite{SILVA}. Since our method involves analyzing {\em distributions} even in sub-domains of the 
Dalitz plot, it  
requires substantial data sets. 

It has been estimated that LHCb will collect very sizable data sets of three-body decays already 
in one nominal LHC year: 
\begin{itemize}
\item
about $10^6$ singly Cabibbo suppressed $D^{\pm} \to \pi^{\pm}\pi^+\pi^-$ and $10^5$ doubly 
Cabibbo suppressed 
$D^{\pm} \to K^{\pm}\pi ^+\pi^-/K^{\pm}K^+K^-$; 
\item
around $10^5$ $B^{\pm} \to K^{\pm}\pi^+ \pi^-$, $B^{\pm} \to K^{\pm} K^+ K^-$ and 
$B^{\pm} \to \pi^{\pm}\pi^+ \pi^-$; 
\item
more than $10^4$ $B^{\pm} \to \pi^{\pm} K^+ K^-$ and the baryonic modes 
$B^{\pm} \to K^{\pm} p \bar p$,  $ \pi^{\pm} p \bar p$. 

\end{itemize} 

The paper will be organized as follows: in Sect. \ref{BASICSVIR} we will briefly review the basics of Dalitz plot analyses, 
introduce a novel observable for probing \cp~symmetry there and comment on the 
isobar model; in Sect. \ref{BDEC} we apply it to 
$B$ mesons decaying into three light mesons and present Monte Carlo studies; in Sect. \ref{DDEC} we discuss analogous 
$D$ decays where one has to face rather different challenges; in Sect. \ref{THTOOLS} we present expectations about in 
which direction and to which degree relevant theoretical tools might get refined in the near future before 
summarizing and giving an outlook in Sect. \ref{OUT}.

\section{Basics and Virtues of the Dalitz Plot}
\label{BASICSVIR}

\subsection{Basics}
\label{BASICS1}

It is of course a mathematical triviality that {\em local} asymmetries are bound to be larger than 
fully integrated ones. Yet a Dalitz plot description translates such a general qualitative statement 
into a much more concrete one. For it exhibits all that can be learnt directly from the data on final states of three 
stable particles and their dynamics. Since the phase space density of the Dalitz plot is constant, any
observed structure reflects the dynamics of the decay.
Enhanced populations in certain mass regions can 
reveal the presence of a strong resonance and indicate their widths. The angular distributions 
characteristic for the spin of the resonances modulate the mass bands.
Distorted or twisted mass bands point to the interference between resonances.  
These observations can be cast into a quantitative 
treatment by making an ansatz for the final state amplitude consisting of terms describing 
the moduli and complex phases of the contributing resonances and the non-resonant contribution. 
These entities contain a great deal of subtle dynamical information. Comparing them for 
\cp~conjugate transitions provides a very powerful probe of \cp~invariance. 
While \cp~violation has to enter through complex phases on the {\em fundamental} level of the underlying dynamics, 
it can manifest itself in the Dalitz plot through differences in both the aforementioned moduli 
of the hadronic resonances and their phases for conjugate transitions. Since there are typically 
several resonances contributing to a decay, ample opportunities arise for \cp~violation to surface in a 
Dalitz plot. Hadronic `complexities' thus represent good news for the {\em observability} of 
\cp~asymmetries. They become a challenge only, when one undertakes to {\em interpret} a signal in a 
{\em quantitative} way. Yet even there a Dalitz analysis provides essential assistance: the reliability 
of Dalitz plot parametrizations can be inferred from the amount of {\em over}constraints they manage to 
satisfy.

However Dalitz studies still retain a measure of model dependance 
due to the choices one makes concerning the resonances to be included and their parameterization 
and also due to the treatment of the 
non-resonant contribution; the S-wave is the largest source of systematics due to strong
dynamics. The greatly different phase space available in $B$ and $D$ decays 
makes for an almost qualitative difference in how to treat them.  We will comment on it later.

While we maintain that such model dependencies can be reduced 
considerably with increasing data sets and, more important, with future theoretical insights, we want to propose a novel 
method for searching for \cp~asymmetries in three-body final states that is robust in two respects: 
it requires no model assumptions and provides an effective filter against effects due to 
statistical fluctuations. Yet first we will make a few rather  technical remarks on how various 
phases enter the interference between neighbouring resonances. 

\subsection{Phases with Breit-Wigner Resonances}

Due to \cpt~invariance \cp~violation can express itself only via a complex and presumably weak phase. For it to become 
observable, we need the interference between two different, yet still coherent amplitudes. Oscillations can provide 
such a scenario -- as can hadronization in general. The latter case 
is usually expressed by stating that the two amplitudes have to exhibit different weak as well as strong 
phases: 
${\cal M} = e^{i\phi^{we}_1}e^{i\delta^{st}_1(f)}|{\cal M}_1| + 
e^{i\phi^{we}_2}e^{i\delta^{st}_2(f)}|{\cal M}_2|$, 
with $(\phi^{we}_1,\delta^{st}_1(f))\neq (\phi^{we}_2,\delta^{st}_2(f))$. We can write it also in terms of 
phases that combine the weak and strong phases: 
\bea
{\cal M} &=& e^{i\delta _1(f)}|{\cal M}_1| + 
e^{i\delta_2(f)}|{\cal M}_2|\; , \; \; \delta_i(f) \equiv \delta^{st}_i(f) +  \phi^{we}_i \\
\overline {\cal  M} &=& e^{i\bar \delta _1(f)}|{\cal M}_1| + 
e^{i\bar \delta_2(f)}|{\cal M}_2| \; , \; \; \bar \delta_i(f) \equiv  \delta^{st}_i(f) - \phi^{we}_i 
\eea
It is often implied that the strong 
phases $\delta^{st}_i(f)$ carry a fixed value for a given final state $f$. This does not need 
to be true. More specifically it will definitely not hold when the final state contains 
a resonance.  The Breit-Wigner excitation curve for a resonance $R$ reads 
\beq 
F^{\rm BW}_{R} (s) = \frac{1}{m^2_{R} - s - i m_{R}\Gamma_{R}(s)}  ~,
\label{BW}
\eeq 
introducing a sizable phase as expressed through 
\beq 
{\rm Im}F^{\rm BW}_{R} (s) = \frac{m_{R}\Gamma_{R}(s)}{(m^2_{R} - s)^2+
(m_{R}\Gamma_{R}(s))^2 } \; , 
\eeq 
where $\Gamma_{R}(s)$ denotes the energy dependent relativistic width. 
In our discussion of $B^{\pm}\to K^{\pm}\pi^+\pi^-$ we will focus on the interference between 
$\rho^0$ and $f_0$ to generate a \cp~asymmetry. The relevant amplitude components for 
$B^+$ and $B^-$ are: 
\begin{eqnarray}
{\cal M}_+ &=& a^{\rho}_+ e^{i \delta^{\rho}_+}F^{\rm BW}_{\rho} \cos \theta + 
a^{f}_+ e^{i \delta^{f}_+}F^{\rm BW}_{f} 
\label{RHOF0AMP1}\\
{\cal M}_- &=& a^{\rho}_- e^{i \delta^{\rho}_-}F^{\rm BW}_{\rho} \cos \theta + 
a^{f}_- e^{i \delta^{f}_-}F^{\rm BW}_{f} 
\label{RHOF0AMP2}
\end{eqnarray}
The $\delta _{\pm}$ contain both the fixed weak and the strong phases with the Breit-Wigner 
functions $F^{\rm BW}$ introducing additional mass dependent strong phases as sketched above. 
For the $f_0$ we have followed BaBar's treatment \cite{BaBark2pi} using the Flatt\`e representation, which reflects the proximity of the $K \bar K$ threshold 
and the ensuing distortion of the resonance curve. 
In Eqs.(\ref{RHOF0AMP1},\ref{RHOF0AMP2}) above $\theta$ is the angle  between the $\pi^-$ and the $K^+$ momenta, measured 
in the $\rho$ rest frame. This angle describes the  angular distribution of a vector meson.After taking the modulus square of these amplitudes one reads off that a \cp~asymmetry will 
arise, when there are non-zero {\em weak} phases. 

Since charm decays proceed in an environment of virulent final state interactions, an  
absence of strong phase shifts in $D$ decays is the least of our concern, since it would happen only 
`accidentally'. Yet in the presence of hadronic resonances it becomes even a `mute' point, 
since the resonance provides a mass dependent strong phase that is predictable in most cases and 
thus actually helps to validate a signal. Resonances then create the more favourable 
scenario. 

Ideally we would apply the method proposed by us (see below) to real primary data. Unfortunately 
we do not have access to those. Therefore we start out by using models that are  
{\em consistent} with {\em existing} data to create a Monte Carlo Dalitz plot;  for 
$B^{\pm}\to K^{\pm}\pi^+\pi^-$ we have been thus `inspired' by Babar \cite{BaBark2pi} and 
for $D^{\pm}\to \pi^{\pm}\pi^+\pi^-$ by E791\cite{E791}. Then we create by hand a single `seed' for a 
\cp~asymmetry and analyze whether our method can uncover it; subsequently we vary 
that single seed. 

We will employ the {\em isobar} model \cite{ISO} for constructing Dalitz plots. 
The amplitude for resonant sub-processes is expressed through Breit-Wigner functions multiplied 
by angular distributions as determined by their spins. The amplitudes of all contributing 
sub-processes are combined coherently with complex coefficients. The latter represent free 
parameters that are fixed from the data using a maximum likelihood fit: the magnitudes of the 
complex coefficients are related to the  fractional contributions of each sub-channel and their 
relative phases reflect the final state interactions between the resonances and the `bachelor' particles. 
In Eqs. (\ref{RHOF0AMP1},\ref{RHOF0AMP2}) we have exemplified the general procedure by writing 
down the amplitude for $B^{\pm} \to K^{\pm} \rho^0/K^{\pm} f_0$. 
These relative phases are treated as constant, since they depend only on the total mass of the system, which in 
this case amounts to the mass of the decaying heavy meson. 
The non-resonant three-body 
contribution is usually assumed to be flat over the Dalitz plot or at least described by a smooth distribution. 

\subsection{The Novel Proposal}
\label{PROP}

The challenge we have to deal with in comparing Dalitz plot populations is one of unbiased pattern recognition. It is thus 
analogous to one faced routinely by astronomers: they often search for something they do not quite 
know what it is -- at least initially -- at a priori unknown locations and having to deal with background sources that are 
all too often not really understood. This sounds like a hopeless proposition, yet 
astronomers have been successful in overcoming these odds. Thus we should be eager to learn from them. 

The Pierre Auger observatory has already adopted the same method for statistical weighting in their 
searches for cosmic ray sources, and we propose to follow suit in defining a search strategy for 
\cp~asymmetries in Dalitz analyses: rather than study the customary {\em fractional} asymmetry 
\beq
\Delta (i) \equiv \frac{N(i) - \bar N(i)}{N(i) + \bar N(i)}
\label{DELTA}
\eeq
in particle vs. anti-particle populations $N(i)$ and $\bar N(i)$ 
for each bin $i$, respectively, one should analyze the {\em significance} 
\beq
^{\mathrm{Dp}}S_{CP} \equiv \frac{N(i)- \bar N(i)}{\sqrt{N(i) + \bar N(i)} } \; , 
\label{SIGMA}
\eeq
which amounts to a standard deviation for a Poissonian distribution
\footnote{We will refer to analyzing $^{\mathrm{Dp}}$S$_{CP}$ instead of $\Delta (i)$ as adopting the `Miranda' procedure 
or as `mirandizing' the \cp~search.}.
We will demonstrate below through 
Monte Carlo studies of $D$ and $B$ decays that analysis of the significance $\sigma$ provides a more robust probe 
of \cp~symmetry.  We will illustrate how the observable $^{\mathrm{Dp}}$S$_{CP}$ is 
highly effective in filtering out genuine asymmetries from statistical fluctuations. 

A final technical comment concerning binning size: in the studies presented below we have 
required bins to contain at least twenty events. This number appears `reasonable', but is somewhat 
ad-hoc. Applying our method to real primary data in the future should shed light on the appropriateness 
of this lower bound.

\subsection{First Summary of the Advantages of Our Proposal}
\label{VIRT}

Analyses of Dalitz plots have so far not `bagged' any success in 
{\em establishing} \cp~violation. Even so we 
expect them to become central probes of \cp~invariance due to the following features: 
\begin{itemize}
\item 
Local asymmetries are bound to be larger than integrated ones thus facilitating the task of controlling 
systematic uncertainties. 
\item 
The latter -- either due to production asymmetries or to detection inefficiencies -- can be probed and controlled 
through the analysis of ratios of particle yields. 
\item 
The bin observable $^{\mathrm{Dp}}$S$_{CP}$ defined in Eq.(\ref{SIGMA}) does not suffer from any model dependance 
and allows a robust search for asymmetries that are small or in relatively small samples. 
\item 
This procedure does not represent a diversion on the (long) path to the ultimate goal, namely 
to arrive at a complete Dalitz plot description and all it can teach us. On the contrary -- it will accelerate 
our progress on that journey providing us with increasingly powerful pointers for where to focus our 
attention and constraints for the Dalitz parametrizations.

\end{itemize}
In the following we will present case studies of $B$ and then $D$ decays to illustrate the general method. 

\section{$B$ Decays}
\label{BDEC}

\subsection{General Remarks}

Decays $B \to h_1h_2h_3$ with $h_i = \pi$, $K$ exhibit a pattern in their Dalitz plots that at first sight 
might look surprising: the bands near the edges are crowded while the interior is sparsely populated. Yet on second thought 
this is as expected. For the phase space available in $B$ decays is quite large, in 
particular for non-charm final states. Those will typically consist of significantly more than three 
stable mesons.  For three meson final states the two primary $\bar qq$ clusters produced in the 
$B$ decays have to recede from each other quickly with untypically low masses; thus they generate the pattern sketched above.  

That final states consisting of just two or three pseudoscalar mesons are a rather untypical 
subset of nonleptonic $B$ decays can be seen also in another way: it has been firmly established that the lifetime of charged 
$B$ mesons exceeds that of neutral ones \cite{PDG}: 
$\tau (B^+)/\tau (B_d) = 1.071 \pm 0.009$ --- 
in agreement with already the first fully inclusive theoretical treatment based on the operator product expansion,  
which traces this difference back mainly to a {\em destructive} interference in nonleptonic $B^+$ 
decays \cite{MIRAGE}. Yet when one sums over the $B\to D\pi$, $B\to D^*\pi$ and $B\to D\rho$ 
channels one finds that there the $B^+$ width exceeds that for $B_d$ by about a factor of two! 
This is in marked contrast to the case of $D$ mesons where the sum of the partial widths for 
$D\to K\pi$, $K^*\pi$ and $K\rho$ already exhibit the same pattern as the total widths.  

There are many modes that carry considerable promise to reveal \cp~violation and shed light on the 
underlying dynamics. We will focus on just one $B$ (and later on just one $D$) mode in this note for two reasons: 
the pedagogical one that we do not want to `over-feed' the reader; and the very practical one that so far little 
experimental information exists about these $B$ decays. In this spirit we will 
discuss $B^{\pm} \to K^{\pm}\pi^+\pi^-$. 
This channel is predicted to have a large component from the Penguin operator. Since that  
operator is derived from a loop process --- i.e., a pure quantum effect --- it  
represents a wide gateway for New Physics. One should also note that the neutral two-body 
counterpart $B_d \to K^+\pi^-$ has already shown a direct \cp~asymmetry \cite{HFAG}. 


Observing a \cp~asymmetry here is unlikely to be the main challenge -- that role is reserved for the question whether 
an observed signal is generated by CKM forces alone or requires the intervention of New Physics that probably provides 
merely a non-leading contribution. We know of no model-independent way to settle this issue and thus have to rely on 
theoretical treatments that are based 
on more than just basic features of QCD, yet still require model assumptions not (yet) derived from QCD. 

\subsection{$B^{\pm} \to K^{\pm}\pi^+\pi^-$}
\label{BTOK2PI}

We will describe this case in considerable detail, since it commands a relatively large 
branching ratio compared to other charmless final states and there is strong evidence for a direct \cp~asymmetry associated with the 
$B^{\pm} \to K^{\pm}\rho^0(770)$ sub-channel \cite{BaBark2pi,bellek2pi}. 
It also provides a clear illustration of the power of our method. 

The moduli and phases of its  amplitudes are `inspired' by BaBar's results \cite{BaBark2pi}. 
We include five resonant and one non-resonant contribution; the latter is assumed to be flat over the 
Dalitz plot purely for reasons of convenience and the lack of a specific alternative. We analyze two versions each with 
a {\em single seed} of \cp~violation, namely one with a \cp~asymmetry in the overall phase 
for the $\rho^0(770)$ and the other one for the $f_0(980)$. To provide a clear demonstration of our method we start out 
by assuming the phase of the $B^+ \to K^+\rho^0(770)$ relative to $B^- \to K^-\rho^0(770)$
to be large, namely 60$^o$, which is still allowed by the 
data \cite{BaBark2pi}. Then we analyze two cases with a significantly smaller phase difference, 
namely 20$^o$ and 10$^o$, respectively. In the latter two cases neither a visual inspection of the Dalitz plot nor using the 
fractional difference $\Delta (i)$ suffice to establish the resulting \cp~asymmetry. Yet an analysis of the 
significance $^{\mathrm{Dp}}$S$_{CP}$ allows even to locate the origin of the asymmetry in the Dalitz plot.

\subsubsection{Model "$\rho^0$"}

The specifics of this version are listed in Table \ref{cpt1}.  
\begin{table}[!h]
  \begin{center}
\begin{tabular}{|c|c|c|c|c|}\hline
\centering
 mode & $a^+$ & $\delta^+$ & $a^-$ & $\delta^-$\\ 
\hline \hline
{$K^*(890)\pi$} & 1.0   & 0.00  & 1.00  &  0.00 \\ \hline
{$K(1430)\pi$} & 2.1   & 6  & 2.1 & 6 \\ \hline
{$\rho(770)K$} & 0.9  & {\bf-34}  & 0.9 &  {\bf 26}\\ \hline
{$f_0(980) K$} &1.0  & 132  & 1.0 & 132 \\ \hline
{$\chi_c K$} & 0.3 &  -143  &  0.3 & -143 \\ \hline
{$NR$} & 0.6 & -109 & 0.6 & -109\\ 
\hline \hline
\end{tabular}
\end{center}
	\caption{Magnitudes and phases, in degrees, of the amplitudes defining Model "$\rho^0$" for our toy 
	Monte Carlo sample. The difference in the $\rho(770)$ {\em phase} for the $B^+$ and $B^-$ 
	channels provides the only source for a genuine \cp~violation.}
    \label{cpt1}
\end{table}
For diagnostic clarity we pick two sets of amplitudes for $B^+$ and $B^-$ decays, shown in Table 1, that differ in a single parameter only, 
namely the {\em phase} of the $\rho(770)K$ amplitude, while all moduli of the amplitudes are the same. 

With these parameters the signal amplitudes for the $B^+$ and $B^-$ are integrated over the Dalitz plot, yielding a 
direct  \cp~asymmetry of about $3 \times 10^{-3}$. For a sample with 300K $B^+ \to K^+\pi^+\pi^-$ decays this corresponds
to 298K $B^- \to K^-\pi^+\pi^-$ events.
  At first one might think that the fully integrated rate 
can show no difference for the $B^+$ and $B^-$ channels, when the only seed of \cp~violation planted 
into the Monte Carlo model is a difference in the overall phase of the $\rho$ contribution. Yet the 
small direct \cp~asymmetry is due to the interference of the triangle -- pun intended -- of  
$K\rho$, $Kf_0$ and $K^*\pi$ amplitudes. 

\begin{figure}[!ht]
\centering
\includegraphics[scale=.55]{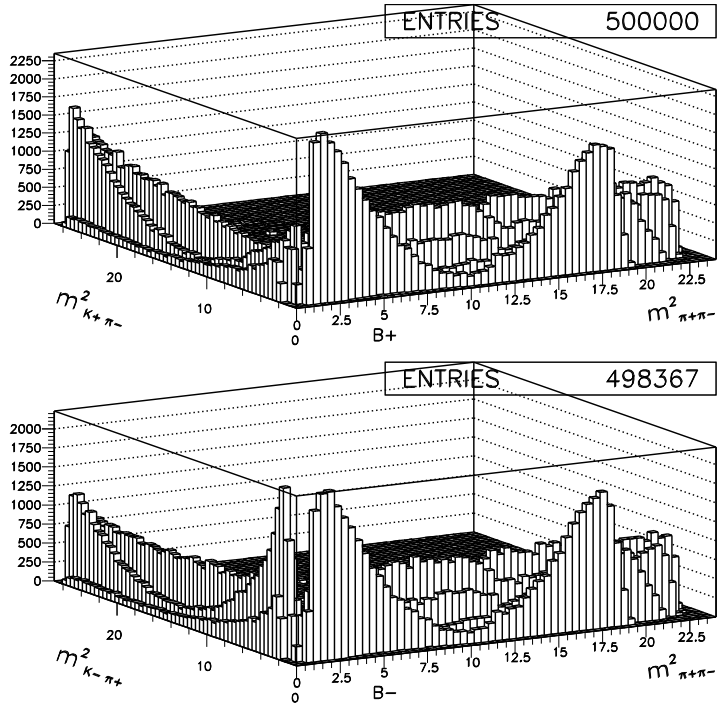}
\caption{ Dalitz plot distribution for $B^+\to K^+\pi^+\pi^-$ (top) and 
$B^-\to K^-\pi^+\pi^-$ (bottom) in Model "$\rho^0$". }
\label{cpf1}
\end{figure}

\begin{figure}[!ht]
\centering
\includegraphics[scale=.55]{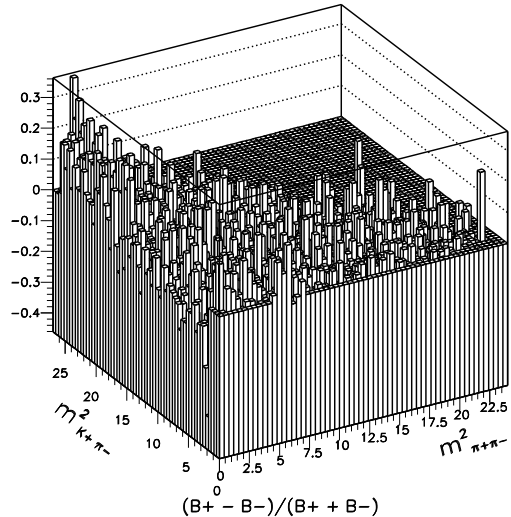}
\caption{ Asymmetry in the Dalitz plot bins for Model "$\rho^0$" as defined by Table \ref{cpt1}. }
\label{cpf1b}
\end{figure}

For the $B^+$ and $B^-$ samples we assume a background 
of about 200K events. The resulting Dalitz plots are shown in Fig.\ref{cpf1}. 
They do not look quite the same. To make their differences more explicit we have plotted the 
fractional asymmetry $\Delta (i)$ of Eq.\ref{DELTA} bin for bin in Fig.\ref{cpf1b}. The resulting 
display is a very noisy one with many bins showing sizable differences, both in the 
$\rho-f_0$ interference region, where our model has to yield a genuine asymmetry, and in the 
central region, where it can{\em not}. 

The `eager' eye might notice that the differences in the former follow a slightly more 
systematic pattern than in the latter, yet it could not be called compelling, in particular if we did not know 
the underlying dynamical structure. 

\begin{figure}[!ht]
\centering
\includegraphics[scale=.55]{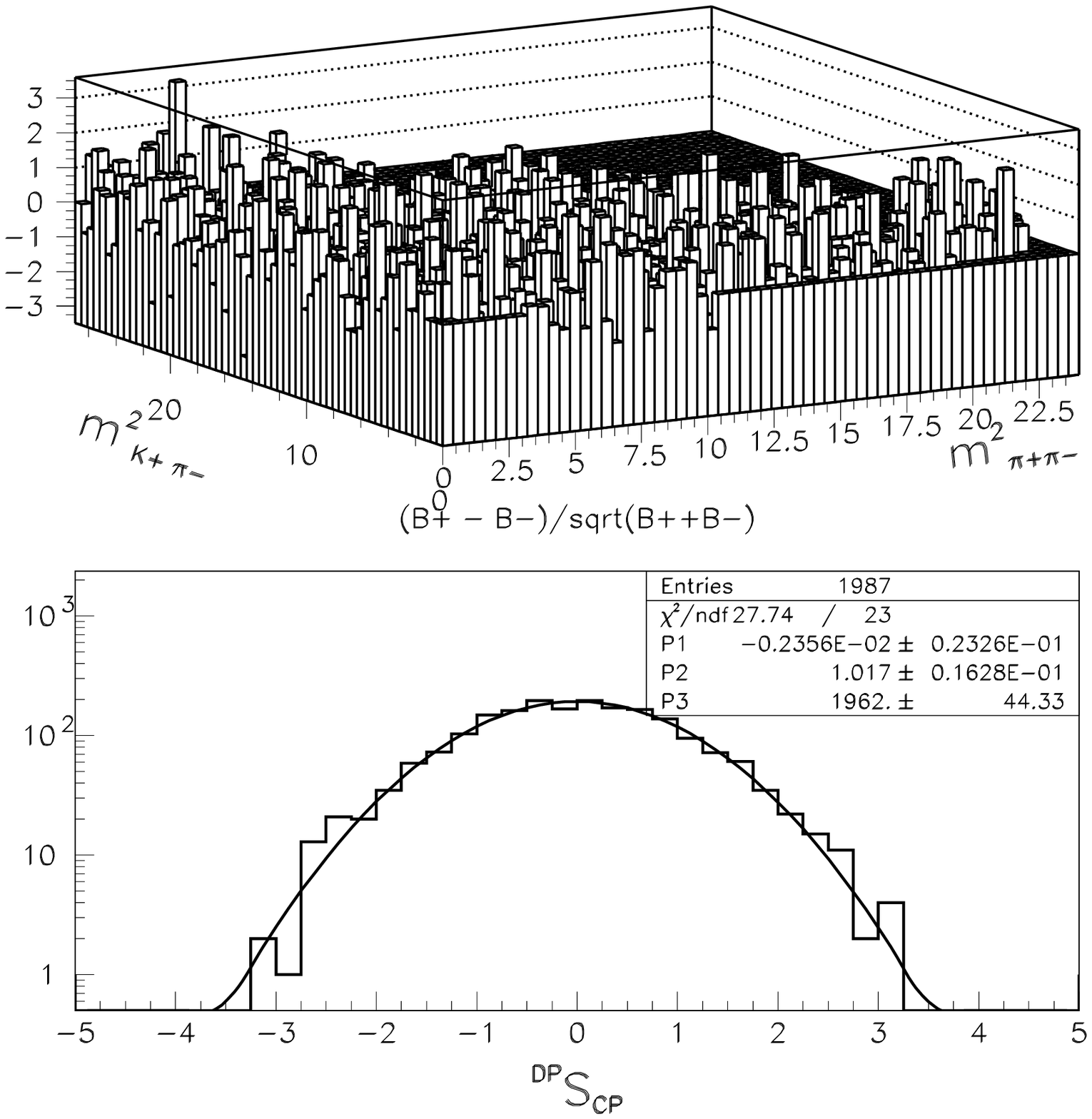}
\caption{Top: Significance $^{\mathrm{Dp}}$S$_{CP}$ plot for two CP conserving 300K signal + 200K
background samples for CP symmetric decays. Bottom: Gaussian fit for the $^{\mathrm{Dp}}$S$_{CP}$ distribution; $P1$, $P2$ and $P3$ 
denote the fit values for the central value, width and normalization parameter, respectively.}
\label{cpf2}
\end{figure}

The effect of the statistical fluctuations can be ilustrated by the following exercise.
We plot in Fig.\ref{cpf2} the significance distribution for a situation where the $B^+$ and $B^-$ Dalitz plots were
generated with exactly the same set of parameters. In this case only statistical fluctuations are observed. 
The upper display shows most bins to exhibit some differences; yet the fact that the 
$^{\mathrm{Dp}}$S$_{CP}$ distribution is completely consistent with a pure Gaussian pattern, as shown in the lower 
display, reveals them to be consistent with mere statistical fluctuations.

\begin{figure}[!ht]
\centering
\includegraphics[scale=.55]{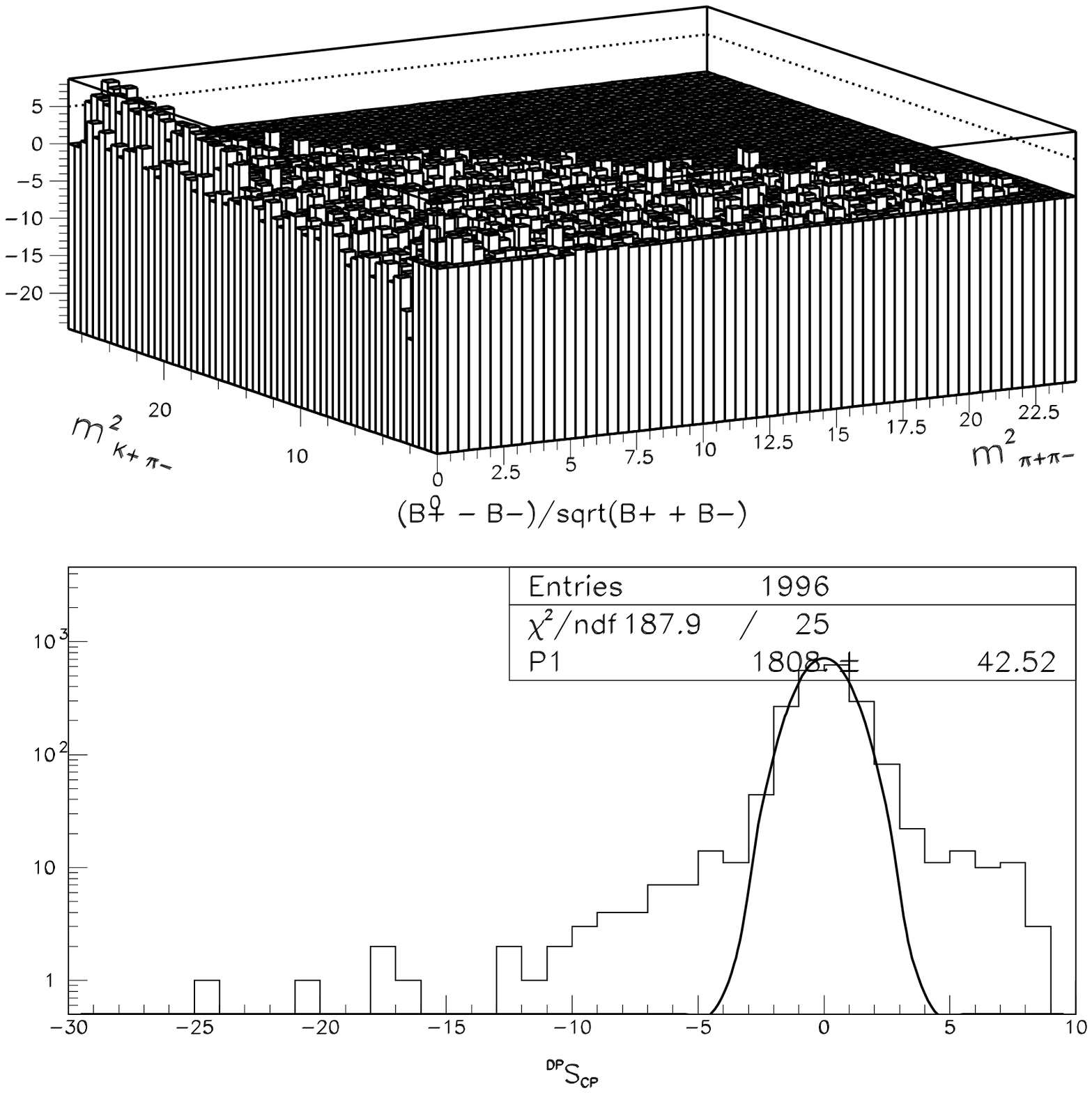}
\caption{ Top: Significance $^{\mathrm{Dp}}$S$_{CP}$ plot for 
$ B^{\pm} \to K^{\pm} \pi^{\mp} \pi^{\pm} $ for model "$\rho^0$". Bottom: $^{\mathrm{Dp}}$S$_{CP}$ for the bins in
Top Figure that pass the statistical cut, fit to a centred Gaussian with unit width. P1 
is the normalization parameter.}
\label{cpf4}
\end{figure}

After our method has successfully passed this null test we return to the model defined by 
Table \ref{cpt1} and Fig \ref{cpf1}.
To obtain a clearer picture we `mirandize' our analysis, i.e. 
turn to the significance $^{\mathrm{Dp}}$S$_{CP}$ defined in Eq.\ref{SIGMA}. 
We plot the resulting values for $^{\mathrm{Dp}}$S$_{CP}$ in the upper display of 
Fig.\ref{cpf4} and its distribution together with a Gaussian fit in the lower display. 

 The Dalitz plot of the significance $^{\mathrm{Dp}}$S$_{CP}$ shows a considerably less noisy pattern 
than before with an obvious asymmetry surfacing in the $\rho$ - $f_0$ interference region. The fact that a genuine 
\cp~asymmetry has surfaced in the Dalitz plot is demonstrated in the lower display: there is no 
acceptable Gaussian fit to the $^{\mathrm{Dp}}$S$_{CP}$ distribution meaning the asymmetries are 
considerable larger than can be generated by statistical fluctuations. 

\begin{figure}[!ht]
\centering
\includegraphics[scale=.35]{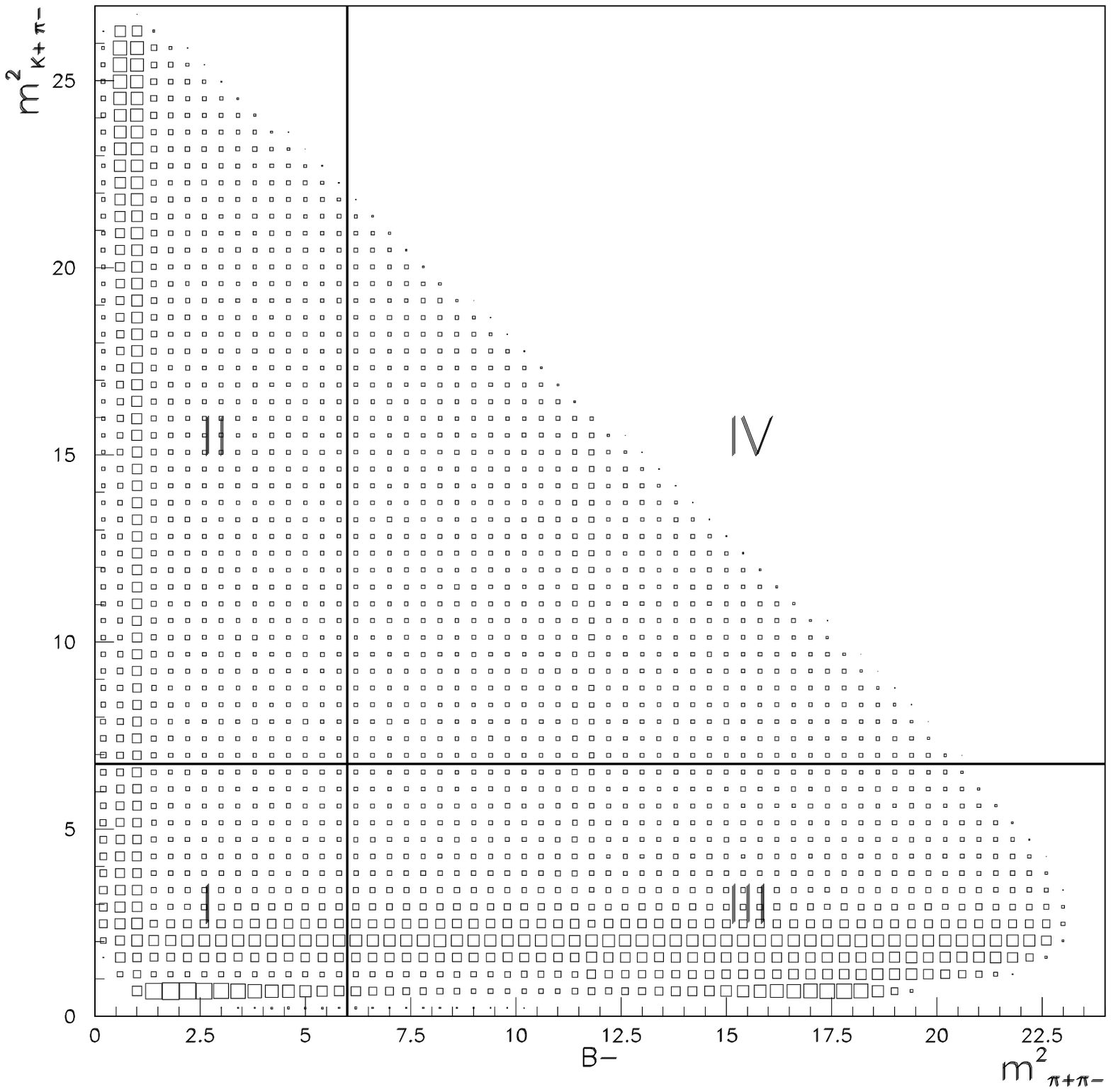}
\caption{ $ B^{\pm} \to K^{\pm} \pi^{\mp} \pi^{\pm} $ Dalitz plot for model "$\rho^0$" divided into
regions.}
\label{cpf3}
\end{figure}

\begin{figure}[!ht]
\centering
\includegraphics[scale=.55]{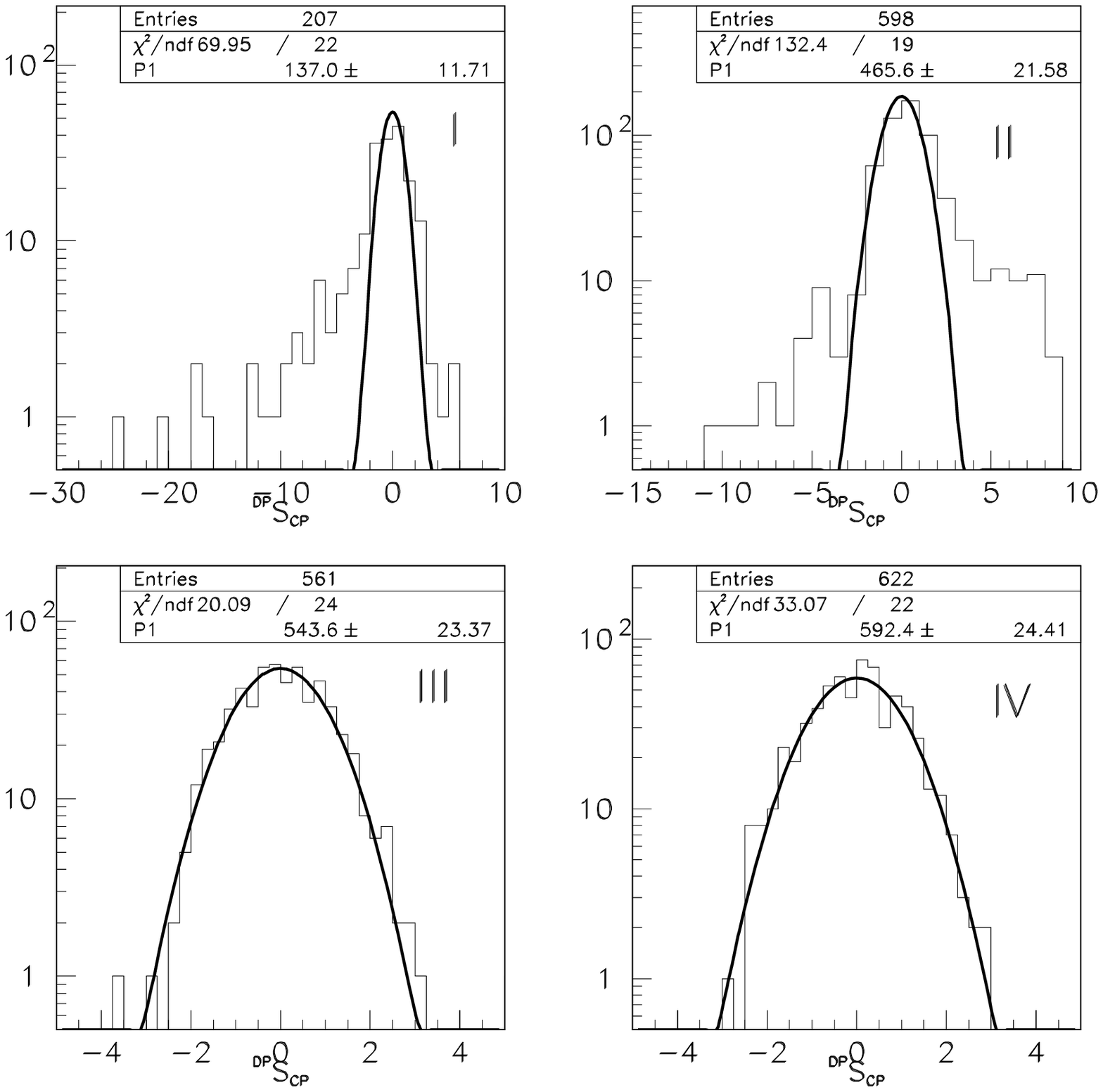}
\caption{ Distribution of figure \ref{cpf4} divided in the regions shown in figure
\ref{cpf3}. P1 is the normalization parameter.}
\label{cpf5}
\end{figure}

Our ambition has of course to go further than just knowing that somewhere in the 
Dalitz plot there is a true \cp~asymmetry -- we want to determine in which subdomain(s) it resides 
and whether it is due to an interference between neighbouring resonances or due to different 
widths of two \cp~conjugate resonances. 
For that diagnosis we divide the Dalitz plot region into subdomains. The choice of 
these subdomains has to be informed by our understanding of the significant subprocesses. In the 
case of $ B^{\pm} \to K^{\pm} \pi^+ \pi^- $ we divide it into the four regions shown in 
Fig.\ref{cpf3}: I and II containing the $\rho (770)$ resonance, III with the 
$K\pi$ resonances and IV populated mainly by background. 
In Fig.\ref{cpf5} we have plotted the $^{\mathrm{Dp}}$S$_{CP}$ distributions separately for these sub-domains. 
The results are very telling: the plot clearly reveals that the asymmetry resides in 
regions I and II, while III and IV show no trace of a genuine \cp~asymmetry -- in full 
agreement with the underlying model chosen to generate these Dalitz plots. 

\subsubsection{Model "$f_0$"}

In this version we use the same model parameters as above (see Table \ref{cpt1}) with two essential differences: 
$\delta ^+ = \delta ^- = -34^o$   ~for the  ~$\rho^0(770)$;   ~$\delta ^+ = 132^o \neq  \delta ^- = 69^o$
for the $f_0(980)$,  
i.e., a phase difference of 63 $^o$. 
Again such a difference is quite compatible with BaBar's findings \cite{BaBark2pi}. 

\begin{figure}[!ht]
\centering
\includegraphics[scale=.55]{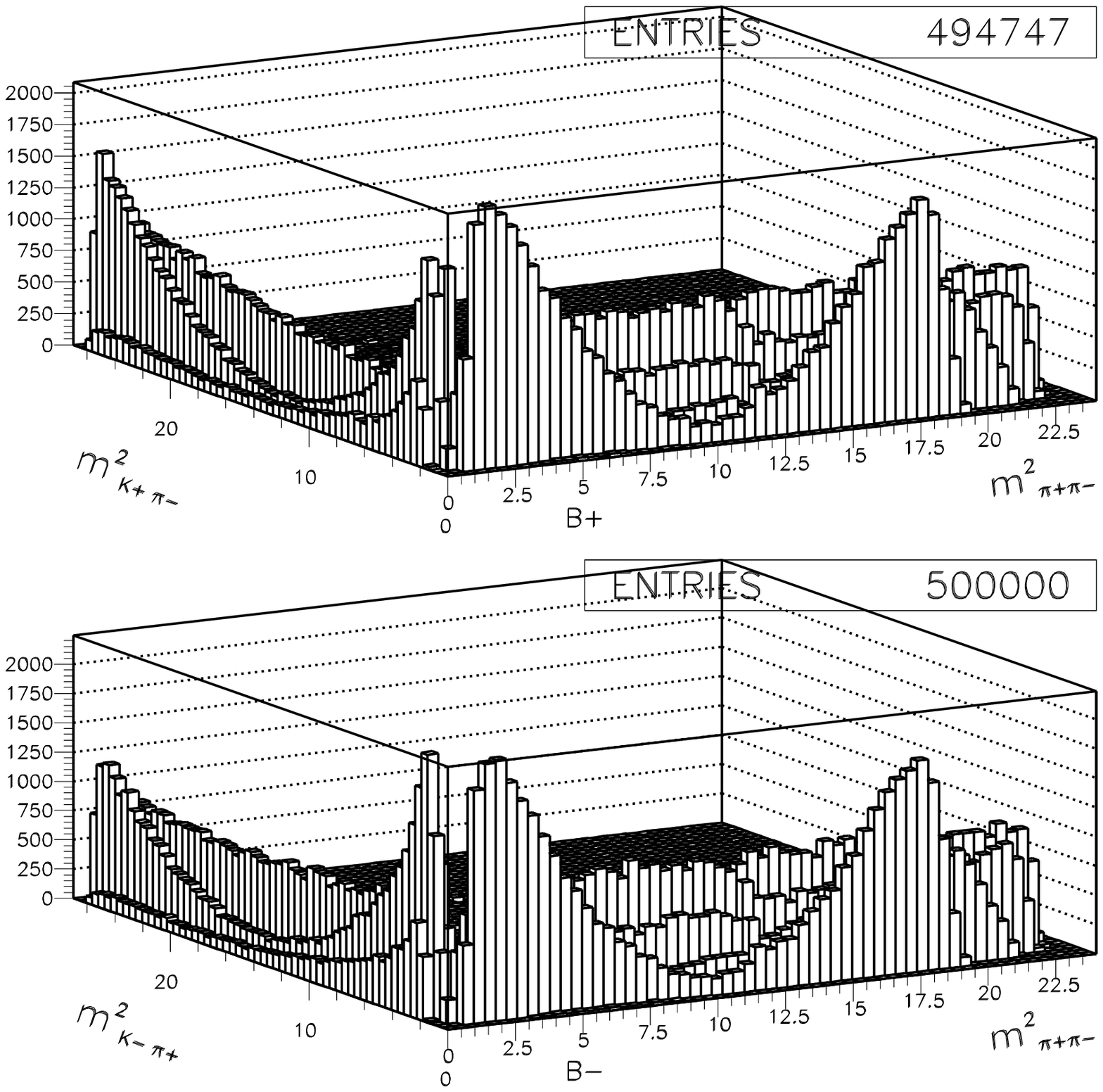}
\caption{ Dalitz plot distributions for $B^+\to K^+\pi^+\pi^-$ (top) and 
$B^-\to K^-\pi^+\pi^-$ (bottom) in Model "$f_0$". }
\label{plf0_1}
\end{figure}

\begin{figure}[!ht]
\centering
\includegraphics[scale=.55]{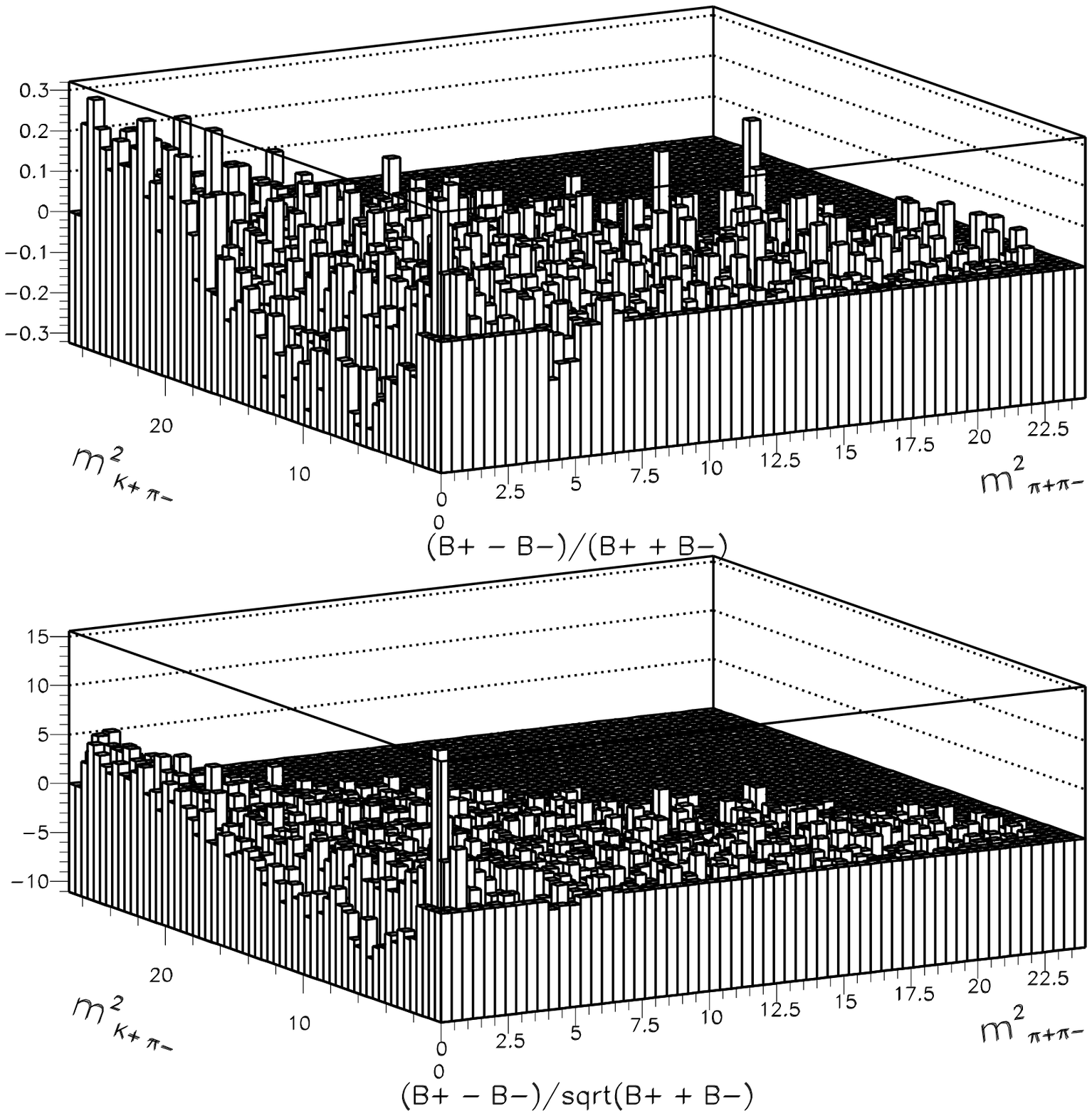}
\caption{ Top: Asymmetry in the Dalitz plot bins for Model "$f_0$". 
Bottom: Plot of the significance $^{\mathrm{Dp}}$S$_{CP}$ for 
$ B^{\pm} \to K^{\pm} \pi^{\mp} \pi^{\pm} $.}
\label{plf0_2}
\end{figure}

\begin{figure}[!ht]
\centering
\includegraphics[scale=.55]{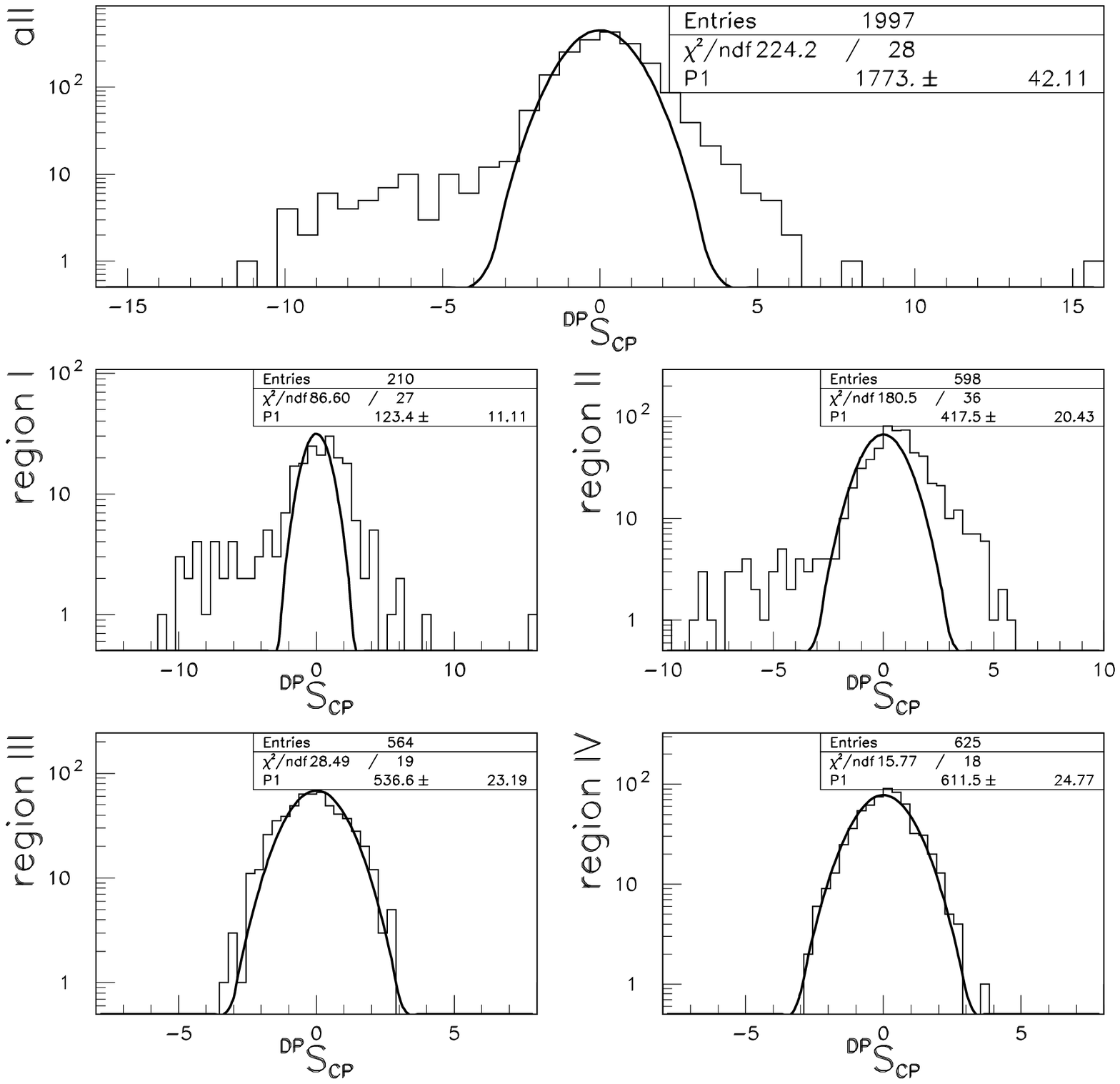}
\caption{ Top row: $^{DP}{\cal S}_{CP}$ for the bins in Fig. \ref{plf0_2}b
 that pass the statistical cut, fit to a centred Gaussian with unit width for model "$f_0$". P1 
is the normalization parameter. Bottom two rows: Distribution of top row divided into the regions shown in 
Fig. \ref{cpf3}. P1 is the normalization parameter.}
\label{plf0_3}
\end{figure}

The $B^+$ and $B^-$ Dalitz plots are shown in Fig. \ref{plf0_1}. 
One can see that the two plots are different. Turning to a plot of the fractional 
asymmetry $\Delta (i)$ shows there are many bin-by-bin asymmetries, yet those exhibit  
again a rather noise pattern, see Fig. \ref{plf0_2}a. 
Once again `mirandizing' the display, i.e., plotting $^{\mathrm{Dp}}$S$_{CP}$ instead of 
$\Delta (i)$, leads to a more organized message, shown in the upper display in Fig. \ref{plf0_2}b. 
In particular, when looking at the $^{\mathrm{Dp}}$S$_{CP}$ distribution of Fig. \ref{plf0_3} we see  
that over and above the statistical fluctuations there is a genuine  \cp~asymmetry.

As before its location can be narrowed down further by dividing the Dalitz plot in the 
four regions of Fig. \ref{cpf3} and plotting the $^{\mathrm{Dp}}$S$_{CP}$ distributions separately for them, see 
Fig. \ref{plf0_3}. It clearly identifies regions I and II as the main origin of the asymmetry. 
That is as it has to be, since the interference between the $K\rho$ and $Kf_0$ amplitudes, which is the "engine" 
of \cp~violation in our model, takes place mainly there.  

\subsubsection{Comparing the "$\rho^0$" and "$f_0$" Models}

The preceding discussion has shown that the $^{\mathrm{Dp}}$S$_{CP}$ observable and its distribution 
provides a powerful tool that in a model independent way allows to establish the existence of 
a genuine \cp~asymmetry over and above statistical fluctuations and even determine the subregion(s) 
of the Dalitz plot, where it originates. For both the two Dalitz models employed above it was mainly the 
$\rho - f_0$ interference domain. 

In addition, a closer analysis allows to distinguish the cases where the asymmetry is driven by a 
difference  in the $K\rho$ and in the $Kf_0$ phase, respectively, for the $B^+$ and $B^-$ decays, see 
Figs. \ref{cpf5} and \ref{plf0_3}b. The discriminator is provided by the interference with the `silent' 
partner, the $K^*\pi$ amplitude. This ability would provide important diagnostics about the underlying dynamics: for 
it would enable us to decide whether the \cp~odd operator generating the asymmetry carries vector or scalar quantum numbers. 

\subsubsection{The case of a smaller $\rho$ phase difference of 20$^o$ and 10$^o$}

\begin{figure}[!ht]
\centering
\includegraphics[scale=.55]{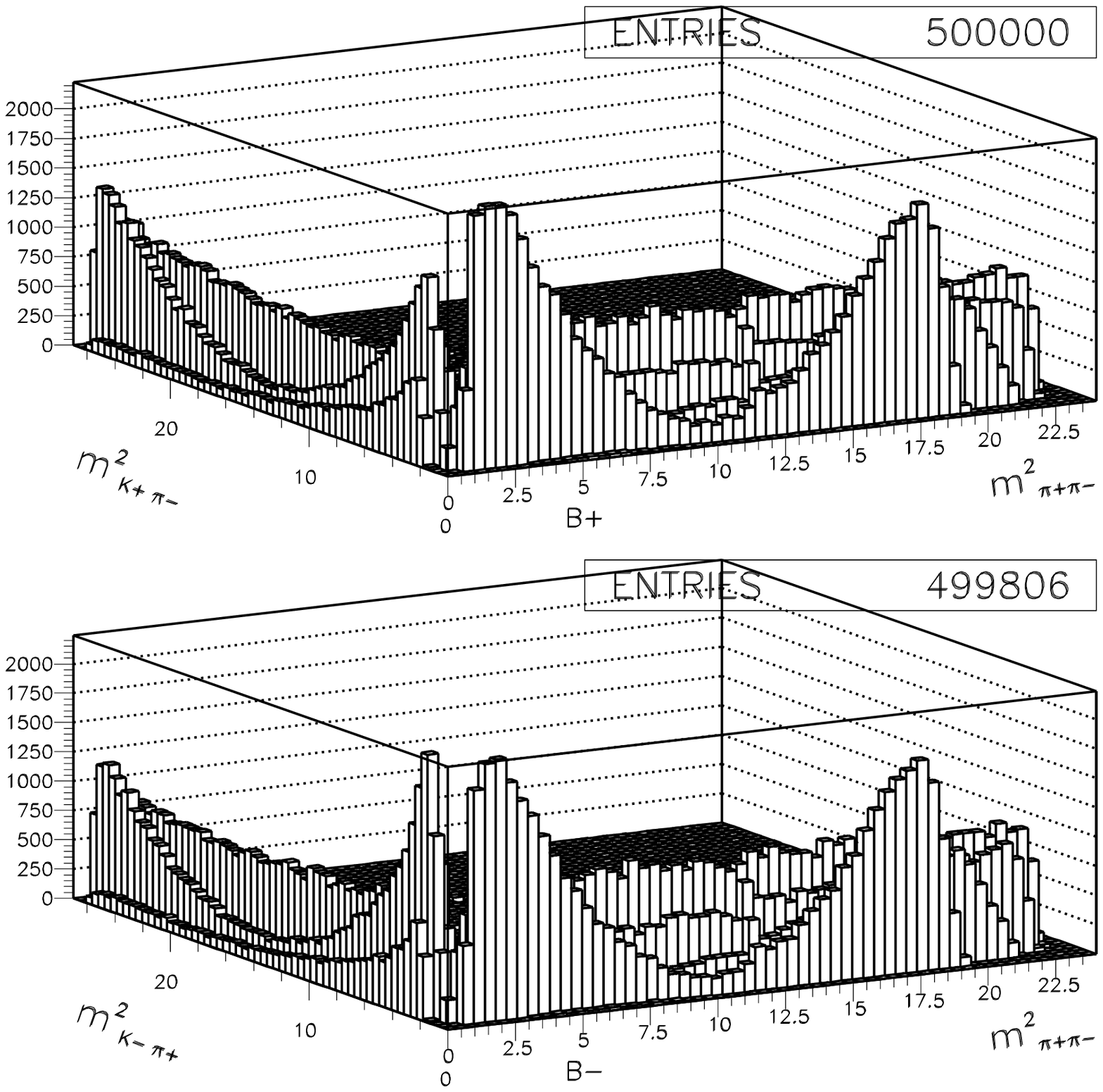}
\caption{ Dalitz plot distributions for $B^+\to K^+\pi^+\pi^-$ (top) and 
$B^-\to K^-\pi^+\pi^-$ (bottom) in a model "$\rho^0$" with a 20$^o$ phase difference. }
\label{plrho20_1}
\end{figure}

\begin{figure}[!ht]
\centering
\includegraphics[scale=.55]{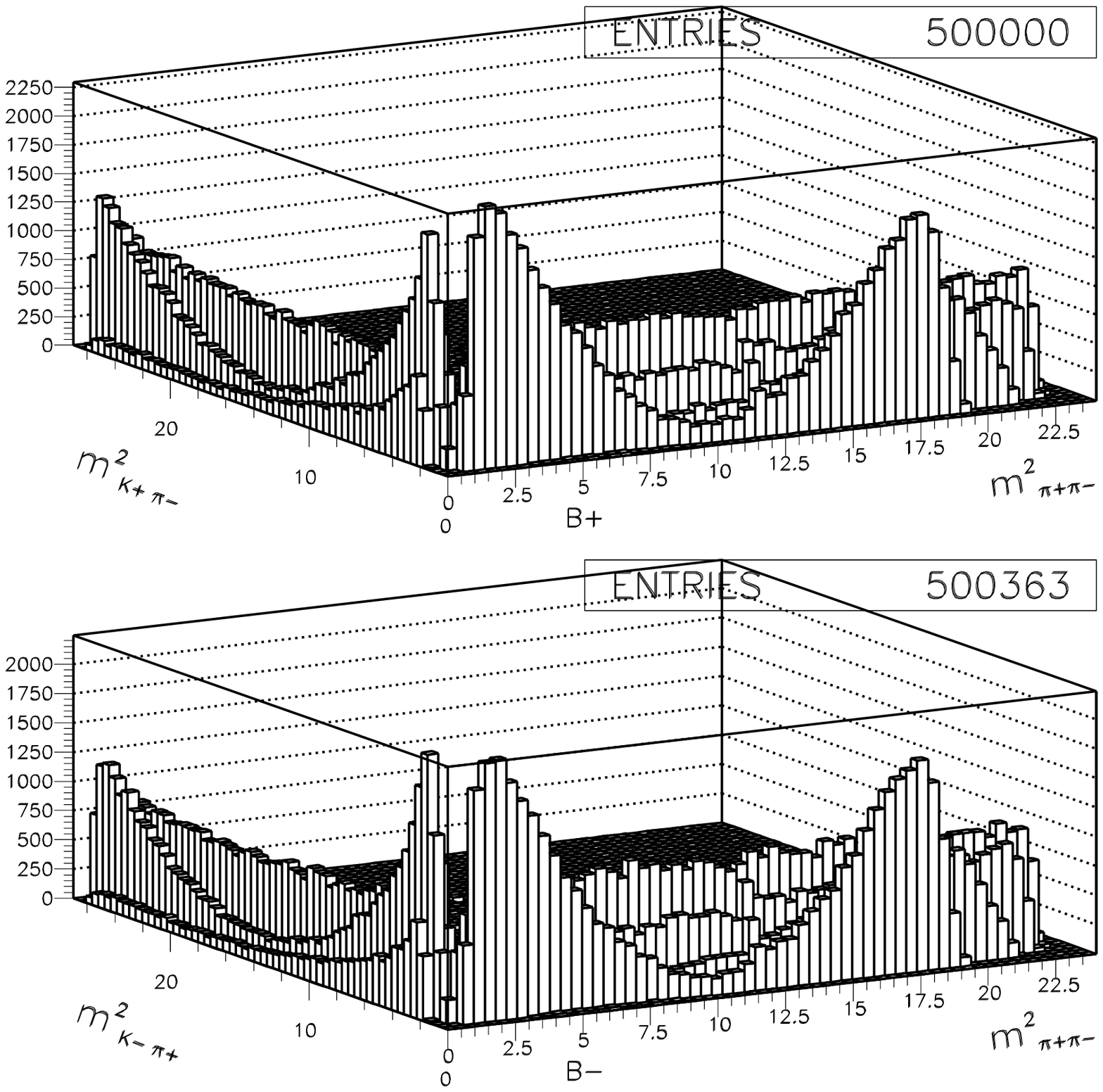}
\caption{ Dalitz plot distributions for $B^+\to K^+\pi^+\pi^-$ (top) and 
$B^-\to K^-\pi^+\pi^-$ (bottom) in a model "$\rho^0$" with a 10$^o$ phase difference. }
\label{plrho10_1}
\end{figure}

\begin{figure}[!ht]
\centering
\includegraphics[scale=.55]{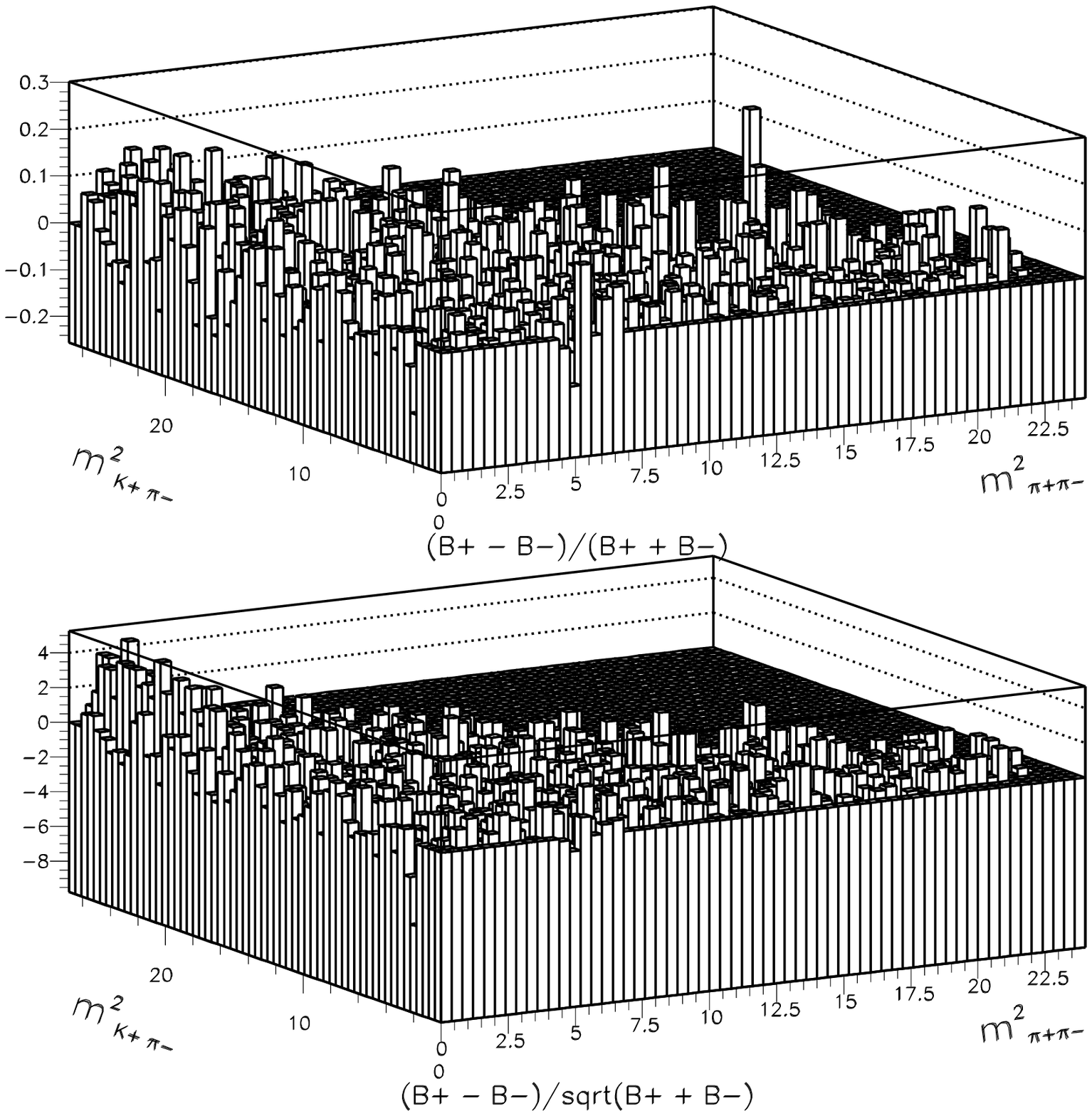}
\caption{Plot of $\Delta (i)$ (top) and 
$^{\mathrm{Dp}}$S$_{CP}$ (bottom)   
for $B^+\to K^+\pi^+\pi^-$ and 
$B^-\to K^-\pi^+\pi^-$ in a model "$\rho^0$" with a 20$^o$ phase difference. }
\label{plrho20_2}
\end{figure}

\begin{figure}[!ht]
\centering
\includegraphics[scale=.55]{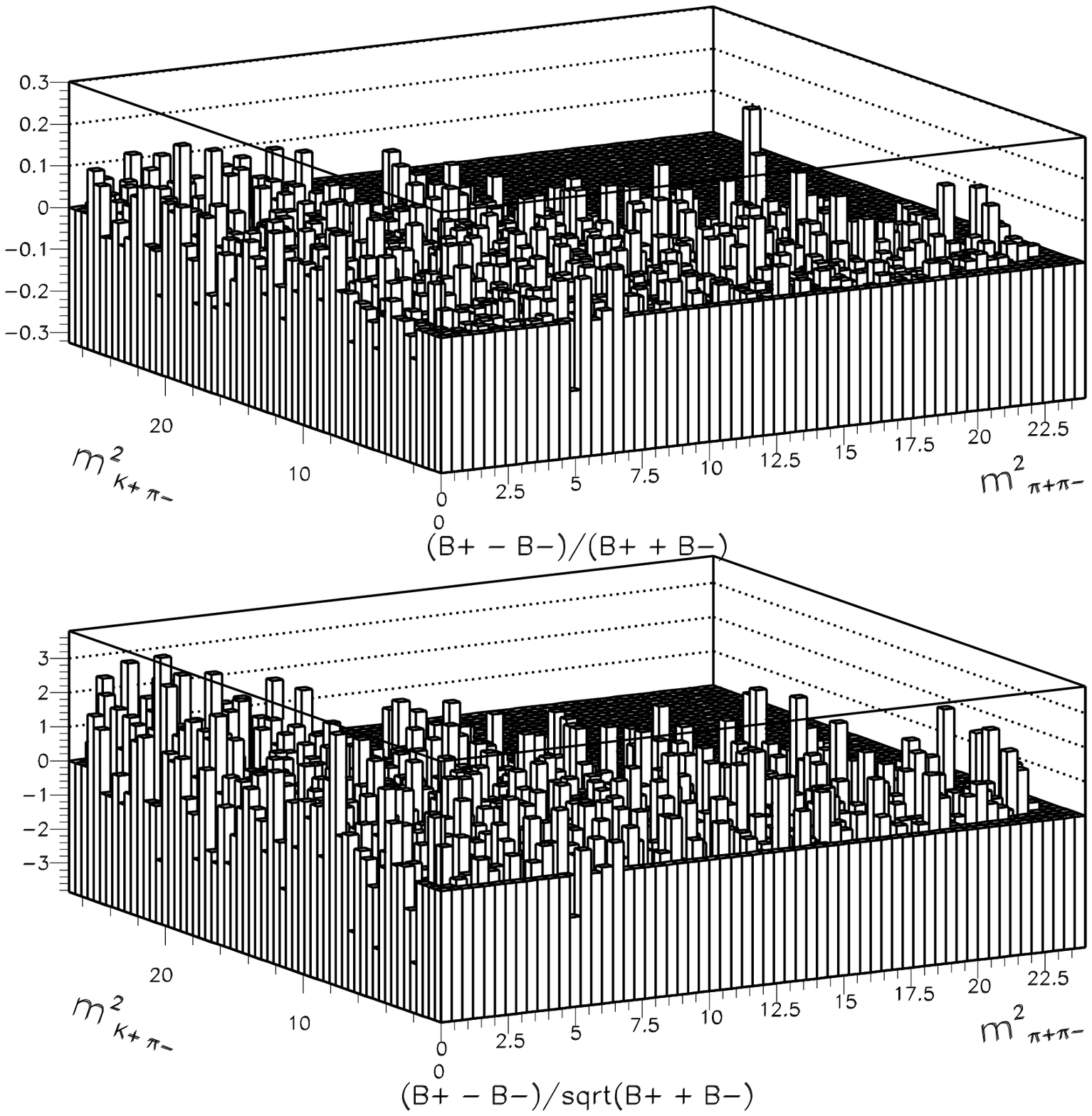}
\caption{Plot of $\Delta (i)$ (top) and 
$^{\mathrm{Dp}}$S$_{CP}$ (bottom) 
for $B^+\to K^+\pi^+\pi^-$ and 
$B^-\to K^-\pi^+\pi^-$ in a model  "$\rho^0$" with a 10$^o$ phase difference. }
\label{plrho10_2}
\end{figure}

\begin{figure}[!ht]
\centering
\includegraphics[scale=.55]{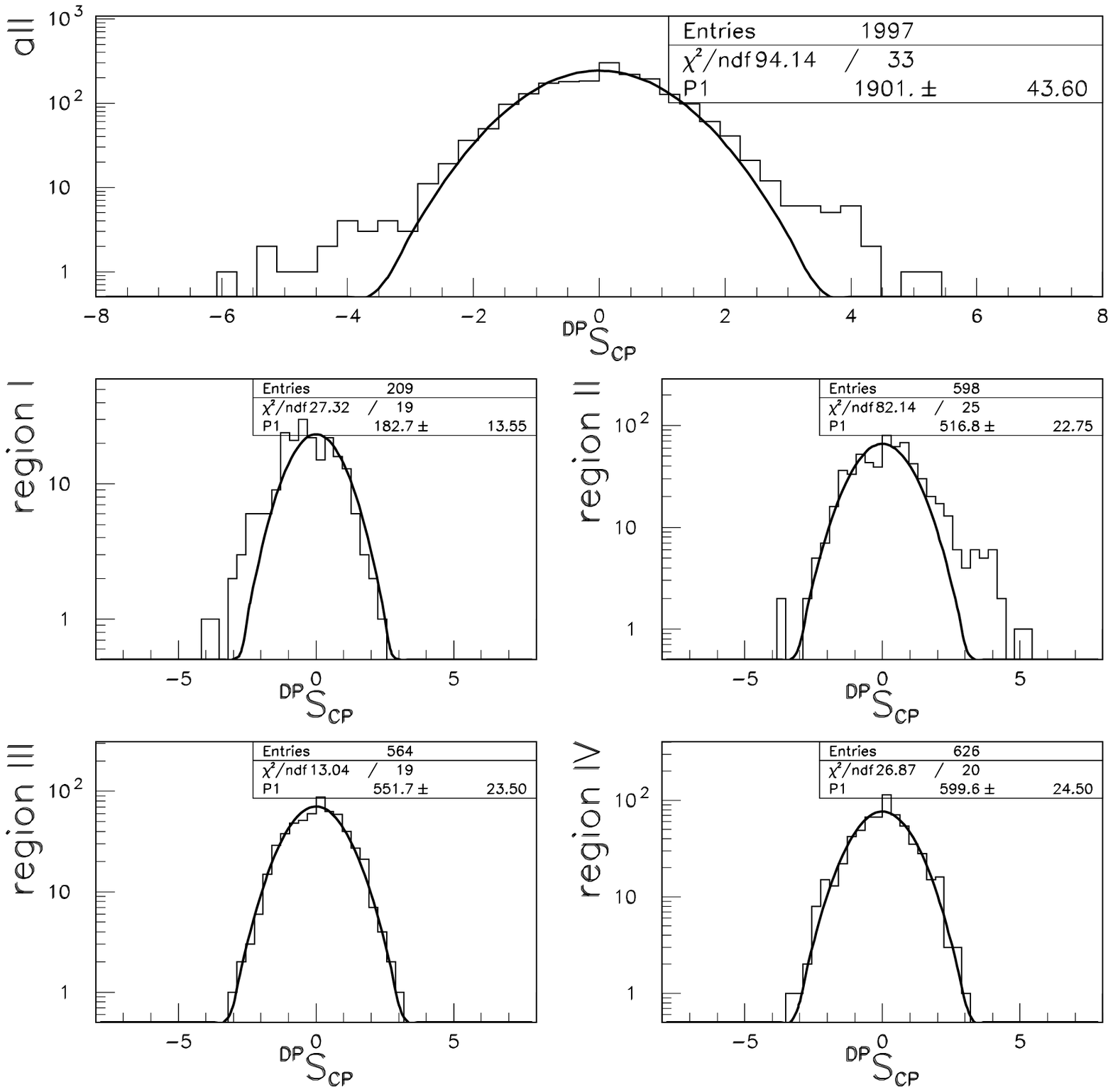}
\caption{ Top row: Distribution of $^{\mathrm{Dp}}$S$_{CP}$ that pass the statistical cut, fit to a centered Gaussian 
with unit width; P1 
is the normalization parameter. Bottom two rows: Distribution of $^{\mathrm{Dp}}$S$_{CP}$ divided into the regions shown in 
Fig. \ref{cpf3} in a model  "$\rho^0$" with a 20$^o$ phase difference. P1 is the normalization parameter. }
\label{plrho20_3}
\end{figure}

The overall phase differences we have assumed for the two models employed above were rather large, although still 
consistent with present data. Consequently even an unsophisticated `look' at the conjugate Dalitz plots suggested 
the existence of a true \cp~asymmetry. Yet for smaller and presumably more realistic values of these phase differences 
one needs the more refined analysis outlined above. 
Instead of the 60$^o$ phase difference in the $\rho^0$ amplitude we had assumed above in our model 
"$\rho^0$", we now assume a phase difference of just  20$^o$ and 10$^o$, respectively, while leaving the other 
parameters as listed in Table \ref{cpt1}. Figs. \ref{plrho20_1} and  \ref{plrho10_1}
show the resulting Dalitz plots separately for the $B^+$ and $B^-$ decays; they look very much the same now.  
In Figs. \ref{plrho20_2} and \ref{plrho10_2} the lower and upper displays show the $\Delta (i)$ and $^{\mathrm{Dp}}$S$_{CP}$ plots,  
for the two scenarios of a 20$^o$ and 10$^o$ phase difference, respectively. 
The lower displays of $\Delta (i)$ are very noisy with no clear message. The upper display 
of $^{\mathrm{Dp}}$S$_{CP}$ shows a systematic 
deviation from zero for the 20$^o$ case, while that can be hardly said for the 10$^o$ case. 
The existence of a genuine asymmetry is demonstrated by the $^{\mathrm{Dp}}$S$_{CP}$ 
{\em distribution} of 
Fig. \ref{plrho20_3}a.

Even better, one can localize the region of origin for the \cp~asymmetry as the one where 
$\rho^0$ and $f_0$ interference takes place, see Fig. \ref{plrho20_3}b.

\begin{figure}[!ht]
\centering
\includegraphics[scale=.55]{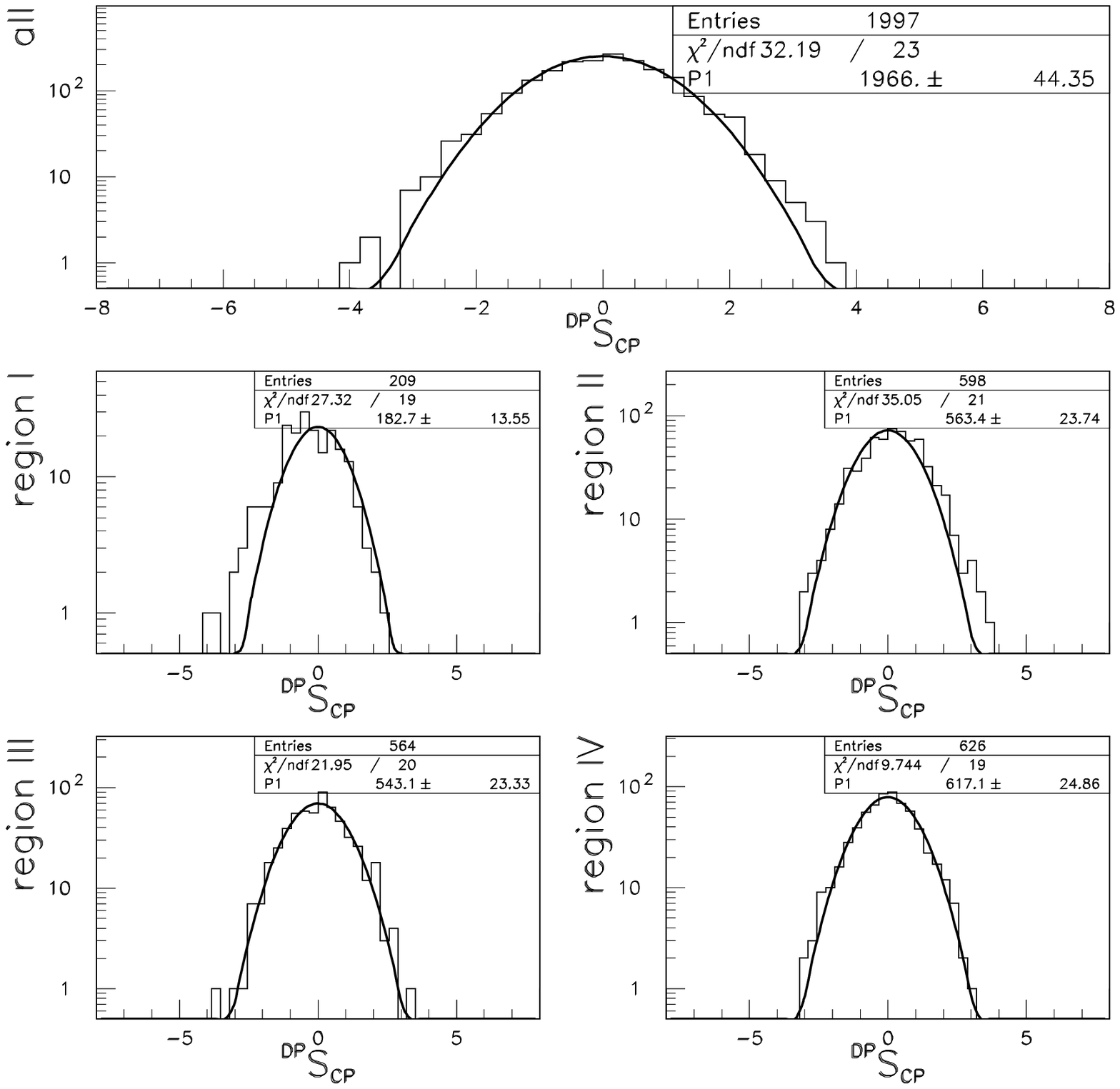}
\caption{ Top row: Distribution of $^{\mathrm{Dp}}$S$_{CP}$ that pass the statistical cut, fit to a centred Gaussian 
with unit width; P1 
is the normalization parameter. Bottom two rows: Distribution of $^{\mathrm{Dp}}$S$_{CP}$ divided into the regions shown in 
Fig. \ref{cpf3} in a model  "$\rho^0$" with a 10$^o$ phase difference. P1 is the normalization parameter.}
\label{plrho10_3}
\end{figure}

The situation for the 10$^o$ case is more delicate. The $^{\mathrm{Dp}}$S$_{CP}$ plot in Fig. \ref{plrho10_3}a 
shows there is no good Gaussian fit to it: the distribution is a bit wider than a Gaussian expression can 
yield, but still symmetric around its maximum. Yet plotting the $^{\mathrm{Dp}}$S$_{CP}$ distributions 
separately for the four regions as before -- Fig. \ref{plrho10_3}b -- reveals a clear message: there is a true 
\cp~asymmetry in regions I and II where $\rho^0$-$f_0$ interference takes place, but none in regions 
III and IV.

We want to stress that for the last two scenarios -- a small phase difference of 20$^o$ and 
10$^o$, respectively, the sophistication provided by an analysis in terms of the 
significance $^{\mathrm{Dp}}$S$_{CP}$ was essential in revealing the underlying dynamics. 

\subsection{Future $B$ Studies}
\label{FUTUREB}

We already mentioned there are several other modes that can be studied with high statistics by LHCb: 
\begin{itemize}
\item 
$B^{\pm} \to \pi^{\pm}\pi^+\pi^-$: like $B^{\pm} \to K^{\pm}\pi^+\pi^-$ it receives contributions from tree as 
well as Penguin operators, yet with the weight of the former enhanced. It thus represents a nicely 
complementary process. 
\item 
The more unconventional channels $B^{\pm} \to \pi^{\pm} p \bar p$, $K^{\pm} p \bar p$ : the presence of the meson
allows us to measure the proton and anti-proton polarization, probing for a \cp~asymmetry, otherwise impossible 
in two-body decays like $B_d \to p \bar p$. 
\item 
$B_d - \bar B_d$ oscillations would lead to Dalitz plots for $B_d \to K_S\pi^+\pi^-$, where the weight 
of different components would shift with the time of decay thus producing time dependent Dalitz plots. 
\item 
The same will happen for $B_s \to K_S K^-\pi^+$, $K_SK^+K^-$, albeit with a much faster 
oscillation rate. 
\end{itemize}
We will address these transitions in future work. 

In this note we have shown how mirandizing the analysis of Dalitz plots -- i.e., studying the `significance' 
distributions -- can act as a powerful filter against statistical fluctuations. Yet real data are also 
vulnerable to systematic experimental uncertainties. For a full validation of our method someone  
has to apply it to real primary data, to which we have at present no access. 

\section{$D$ Decays}
\label{DDEC}

The SM generates a relatively dull weak phenomenology for charm transitions: 
`slow' $D^0 - \bar D^0$ oscillations and tiny \cp~asymmetries; this, however, makes it a 
promising landscape to search for New Physics \cite{CICERONE,BURD,PETROV}. At the same time 
we have to analyze more closely how slow is `slow' quantitatively and how tiny is `tiny'. One has to concede 
that SM dynamics might saturate the observed size of 
  ~$x_D = \Delta M_D/\overline \Gamma_D$   ~and    ~$y_D = \Delta \Gamma_D/\overline \Gamma_D$,  and that 
CKM forces can produce \cp~asymmetries on the ${\cal O}(10^{-3})$ level in singly Cabibbo 
suppressed (SCS) modes. Furthermore, ignoring $D^0 - \bar D^0$ oscillations, purely Cabibbo 
allowed (CA) and doubly suppressed (DCS) channels (i.e. those with{\em out} a $K_S$ or $K_L$) cannot exhibit 
direct \cp~violation. Any such effect in DCS modes and one on the about 0.01 or larger level in 
SCS decays will thus establish the intervention of New Physics. Basing such claims on 
somewhat smaller effects will require theoretical progress that appears quite feasible. 

The phase space available in $D$ decays is significantly smaller than in $B$ decays. Two- and 
quasi-two-body channels make up more than half of the full nonleptonic width. Furthermore the Dalitz plots are 
populated more thorougly than for $B$ decays. 
Determining the impact of individual contributions therefore amounts to a more delicate task.

\subsection{$D^+ \to \pi^+\pi^+\pi^-$}

In the SM there are already two different 
amplitudes contributing to these SCS transitions, and they carry a relative weak phase, 
albeit a tiny one $\sim$ ${\cal O}(\lambda^4) \sim 10^{-3}$. Finding \cp~asymmetries significantly larger than 
$10^{-3}$ would provide strong prima facie evidence for the presence of New Physics. 
Searching down to asymmetries as small as $10^{-3}$ requires huge statistics as well as 
excellent control 
over systematics; the method proposed by us should be a powerful tool in taking up such a challenge. 
Probing the whole Dalitz plot with its various structures should allow us to make the case for New Physics even 
compelling making use also of the anticipated theoretical refinements sketched below. 

It should be noted that New Physics scenarios like the Littlest Higgs Model with T parity could have 
an observable impact here through new 
heavy states appearing as virtual particles in Penguin diagrams \cite{BBBR}. 

We adopt a decay model containing four components, namely 
\begin{itemize}
\item 
$D^{\pm} \to \rho ^0 \pi^{\pm}$, 
\item 
$D^{\pm} \to \sigma ^0 \pi^{\pm}$. 
\item 
$D^{\pm} \to f_0 \pi^{\pm}$
\item
a uniform non-resonant $D^{\pm} \to \pi^+\pi^-\pi^{\pm}$. 
\end{itemize}

We have used  the values obtained by Fermilab experiment E791 \cite{E791} for the magnitudes and phases
of the amplitude coefficients.

A difference in the $\sigma$ phase -- a very conceivable scenario -- will affect many parts of the 
Dalitz plot and induce \cp~asymmetries, since the $\sigma$ possesses a very sizable width relative to 
the phase space available in $D$ decays. The resulting complexities are very intriguing and will be analyzed in a separate paper. 

The case of the $f_0$ amplitude having a different phase in $D^+$ and $D^-$ decays is very interesting for another 
reason: as long as it has any $\bar uu$ or $\bar dd$ 
component it will contribute 
in this final state. We have found that this relatively small contribution can still produce a
clear signature in \cp~asymmetries mainly due to the narrow width of the $f_0$. Details of the required analysis 
will also be given in the future paper. 

In this pilot study we will focus on one scenarios, which leads to clean signatures, namely 
those with a 1\% (equivalent to 3.6$^o$) phase difference in $\rho^0$ amplitude. 
We have selected much smaller phase differences here than for our discussion of $B$ decays above for two reasons: 
(i) They represent much more realistic New Physics scenarios. (ii) Due to the considerably 
smaller phase space and thus shrunk Dalitz plot areas one can expect such effects to be still observable.  

\begin{figure}[!ht]
\centering
\includegraphics[scale=.55]{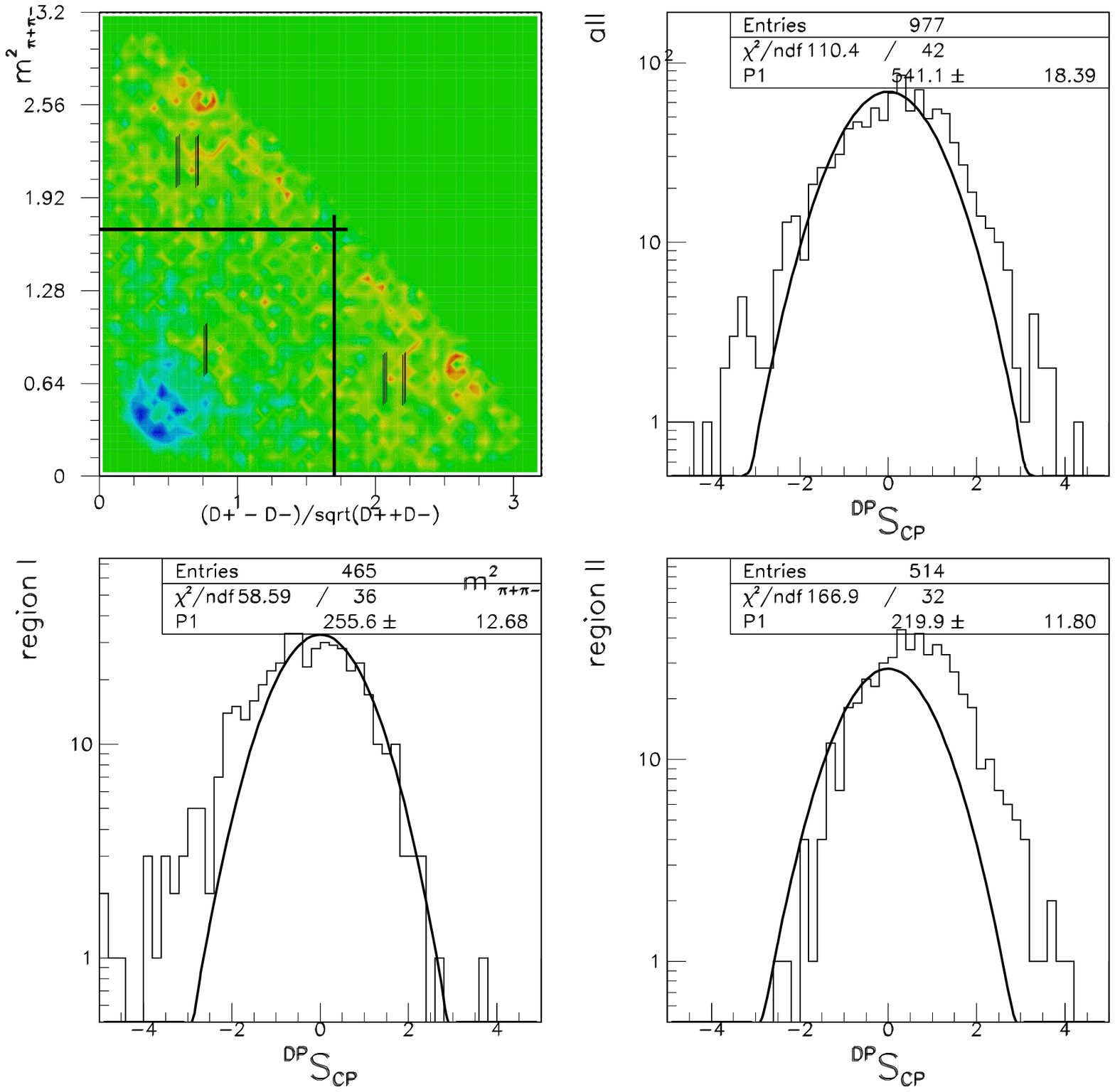}
\caption{ Dalitz plot for $D^{\pm}\to \pi^{\pm}\pi^+\pi^-$ in a model with a 1\% ($3.6^o$) phase difference in the 
$\rho^0$ amplitude with sub-domains I and II; distributions of the significance 
$^{\mathrm{Dp}}$S$_{CP}$
for the whole plot and the two sub-domains I and II; P1 is the normalization parameter.}
\label{Dto3PI}
\end{figure}
In Fig. \ref{Dto3PI} we display the Dalitz plot for this model and the $^{\mathrm{Dp}}$S$_{CP}$ distribution for the whole 
plot as well as for the two regions I and II. The overall $^{\mathrm{Dp}}$S$_{CP}$ distribution unequivocally 
reveals the existence of a \cp~asymmetry, since it does not at all follow a Gaussian fit. The 
distributions for the two regions I and II exhibit a very telling pattern, namely a rather 
asymmetric distribution of the bin-wise \cp~asymmetries. That is how it has to be for the line 
dividing I and II chosen to go through the gap between the two $\rho$ `lobes', as can be seen by
 straightforward arithmetic.  Applying  Eqs. (\ref{RHOF0AMP1},\ref{RHOF0AMP2}) to  $D^{\pm}\to \pi^{\pm}\pi^+\pi^-$,
  with an analogous, though smaller  $\rho$ - $f_0$ interference 
 used in those equations leads to following difference in the  $D^{+}\to \pi^{+}\pi^+\pi^-$
- $D^{-}\to \pi^{-}\pi^+\pi^-$ amplitude squared:

\begin{eqnarray}
\Delta{\cal M} &=&{|\cal M}_+|^2  - |{\cal M}_- |^2  =[ (a^{\rho}_+)^2 - (a^{\rho}_-)^2] |F^{\rm BW}_{\rho}|^2
\cos^2 \theta+ [ (a^{f}_+)^2 - (a^{f}_-)^2] |F^{\rm BW}_{f}|^2\nonumber \\
&&+2\cos\theta |F^{\rm BW}_{\rho}|^2|F^{\rm BW}_{f}|^2 \times \nonumber \\
&&\{[ (m_{\rho}^2 -s) (m_f^2-s) - m_{\rho}\Gamma_{\rho} m_{f}\Gamma_{f}][a^{\rho}_+a^{f}_+ \cos(\delta^{\rho}_+ 
- \delta^{f}_+) - a^{\rho}_-a^{f}_-  \cos(\delta^{\rho}_- - \delta^{f}_-)]\nonumber \\
&&-[m_{\rho}\Gamma_{\rho}(m_f^2-s) - m_{f}\Gamma_{f}(m_{\rho}^2 -s))
[a^{\rho}_+a^{f}_+ \sin(\delta^{\rho}_+ - \delta^{f}_+) - a^{\rho}_-a^{f}_-  \sin(\delta^{\rho}_- - \delta^{f}_-)]\}\nonumber \\
&&\label{DeltaM}
\end{eqnarray}
The term quadratic in $\cos\theta$ is responsible for the parabolic shape of the spin one 
resonance seen in Fig.\ref{cpf1}. Yet the interference generates a term linear in $\cos\theta$.
Therefore the interference is destructive in region I of Fig. \ref{Dto3PI}-- thus implying 
fewer events for $D^+$ than $D^-$ -- and the opposite in region II.

 This example illustrates the power of the mirandizing procedure to unequivocally uncover 
even a small asymmetry and track its local origin in the Dalitz plot.  

\subsection{Future $D$ studies}
\label{FUTD}

As before with $B$ decays many promising channels await careful study: 
\begin{itemize}
\item 
The more complex scenarios in $D^{\pm} \to \pi^{\pm}\pi^+\pi^-$, where the seeds for 
\cp~violation reside in the $f_0$ and $\sigma$ amplitudes deserve detailed analysis. 
\item 
The doubly Cabibbo suppressed modes $D^{\pm} \to K^{\pm}\pi^+\pi^-$, $K^{\pm}K^+K^-$ could reveal a 
new source of direct \cp~violation \cite{DIQUARK}. 
\item 
With the observation of $D^0 - \bar D^0$ oscillations one expects time dependent Dalitz plots to 
emerge in $D^0 \to K_S\pi^+\pi^-$. This time evolution will allow to differentiate between direct and 
indirect \cp~asymmetries. 

\end{itemize}

\section{On Refining the Theoretical Tools}
\label{THTOOLS}

Even lattice QCD does not allow to treat final state interactions as a matter of principle, except for 
kaon decays, where elastic unitarity can be assumed. Elastic unitarity makes little sense for $B$ decays; 
it might be an approximation of some value in $D$ decays, but we have no reliable even 
semi-quantitative estimate for how good an approximation it might be. 

 We should be able to clarify the picture at least somewhat by adopting `theoretical 
engineering' \cite{BESBOOK}: One considers $D_{(s)} \to PP$ ($P$ = pseudoscalar meson) on all Cabibbo levels 
for $D^0$, $D^+$ and $D_s^+$ mesons. Relying on a modicum of theoretical judgement one 
selects diagrams deemed relevant for these processes and expresses their amplitudes in terms of 
the known CKM factors and radiative QCD corrections and the a priori unknown moduli and 
strong phases of their matrix elements. Fitting these expressions to a comprehensive body of well measured
 branching ratios one fixes these moduli and strong phases. The resulting {\em over}constraints  
provide a check on the reliability of such a fit. The analogous procedure is then applied to 
$D_{(s)} \to PV$ ($V$ = pseudoscalar meson). While such an analysis cannot replace a full Dalitz 
plot description, it can provide valuable constraints on the latter. 

Alternatively we should be able to develop some framework where we can have 
a semi-quantitative treatment of the interference of a narrow resonance with a broad non-resonant 
contribution that to first approximation can be considered even as flat.

\section{Summary and Outlook}
\label{OUT}
So far Dalitz plot studies have not established any \cp~violation  with at least five sigma 
significance --- yet we are confident this period will soon come to an end.
We actually expect such studies to become a central tool for obtaining a more detailed picture of and 
perspective on limitations of \cp~invariance. An acceptable description of the 
Dalitz plot usually has to satisfy a sizable number of {\em over}constraints, which provides a powerful 
validation tool to control systematics. Furthermore --- and maybe even more importantly --- 
it provides us with information about the Lorentz structure of the underlying transition operator that 
cannot be inferred from partial rate asymmetries in two-body final states. 

A full fledged Dalitz plot description thus represents the  `holy grail' in our quest for mapping out 
\cp~violation in $B$, $D$ and maybe even top quark decays. The journey there will however require 
a substantial amount of time, as it is with all `holy grails'. It also remains to be seen to which degree there 
will arise uncertainties due to an irreducible model dependance. The method we have proposed in this note 
for searching for \cp~asymmetries in the populations of Dalitz plots is {\em not} meant to replace Dalitz plot parametrizations:  
\begin{itemize}
\item 
The proposed method will allow to establish the existence of \cp~asymmetries with {\em more limited statistics} and identify 
their topography in the Dalitz plot in a {\em robust and model independent} way. 
\item
Furthermore isospin sum rules \cite{ISOSR} can already be applied to its findings. 

\item 
It will speed up the construction of the full Dalitz plot description and provide powerful validation for it. 

\end{itemize}

In this paper we have described a model independent method for establishing the existence of a \cp~asymmetry in a Dalitz plot and inferring its location. To fully gauge its power it is 
important to apply it to high statistics primary data with their experimental systematic uncertainties.  
If it passes that test, then one can study how to extract maximal information about its parameters, in particular the 
weak phase producing it. The relevant expression is given in Eq.(10). Various methods can be employed to achieve such a goal; 
finding the optimal one requires future detailed analysis. \footnote{We thank J. Appel for forcing upon us an illuminating discussion of this point.}
Various features can be employed to discriminate between SM and potential New Physics effects. In $B \to K \pi \pi$ the SM 
can generate a significant weak phase only through its $(V-A)\times (V-A)$ currents, since the $b\to s$ Penguin operator does 
not carry a weak phase. New Physics thus could make its presence felt through producing a weak phase for a scalar state
like the $f_0$.

Clearly a large amount of also theoretical work is required. While we should not count on theorists achieving miracles, 
we can expect a positive learning curve for them.

\vspace{0.5cm}

{\bf Acknowledgments:} This work was supported by the NSF under the grant number PHY-0807959 and by CNPq. 
One of us (IB) thanks the CBPF for the kind hospitality extended to him, while this work was completed. 
We have benefitted from constructive criticism by T. Gershon and the referee. 
We are also grateful to Jo\~ao Torres and Chris Howk for clarifying conversations  about issues related  to astronomy.

\vspace{4mm}

 
\end{document}